\documentclass[final,5p,twocolumn,10pt]{elsarticle}
\usepackage{epstopdf}
\usepackage{multirow}
\usepackage{color}
\usepackage{amssymb,amsmath,bm}
\usepackage{pdflscape}
\usepackage{booktabs}

\newcommand{\ucd}[1]{\stackrel{\nabla}{#1}}
\renewcommand{\vec}[1]{\mathbf{#1}}

\journal{Non-Newtonian Fluid Mechanics}

\biboptions{compress}

\begin{document}

\begin{frontmatter}

\title{Geometric scaling of elastic instabilities in the Taylor-Couette geometry: \\ A theoretical, experimental and numerical study}

\author[sb]{Christof Schaefer}
\address[sb]{Department of Experimental Physics, Saarland University, 66123 Saarbruecken, Germany}

\author[edi]{Alexander Morozov}
\ead{alexander.morozov@ed.ac.uk}
\address[edi]{SUPA, School of Physics and Astronomy, The University of Edinburgh, James Clerk Maxwell Building, Peter Guthrie Tait Road, Edinburgh, EH9 3FD, United Kingdom}

\author[sb]{Christian Wagner}
\ead{c.wagner@mx.uni-saarland.de}

\begin{abstract}
We investigate the curvature-dependence of the visco-elastic Taylor-Couette instability. The radius of curvature is changed over almost a decade and the critical Weissenberg numbers of the first linear instability are determined. Experiments are performed with a variety of polymer solutions and the scaling of the critical Weissenberg number with the curvature against the prediction of the Pakdel-McKinley criterion is assessed. We revisit the linear stability analysis based on the Oldroyd-B model and find, surprisingly, that the experimentally observed scaling is not as clearly recovered. We extend the constitutive equation to a two-mode model by incorporating the PTT model into our analysis to reproduce the rheological behaviour of our fluid, but still find no agreement between the linear stability analysis and experiments. We also demonstrate that that conclusion is not altered by the presence of inertia or viscous heating. The Pakdel-McKinley criterion, on the other hand, shows a very good agreement with the data.
\end{abstract}


\end{frontmatter}

\section{Introduction}

Simple flows of fluids are often unstable though the mechanism of instability is dependent on the type of the fluid. In Newtonian fluids, flow instabilities and the transition to turbulence are driven by inertia \cite{Chandrasekhar1961}, while in complex fluids the instabilities can be caused by both inertia and anisotropic elastic stresses \cite{Larson1992,Shaqfeh1996,Morozov2007}. One of the best studied classes of complex fluids are dilute polymer solutions that are formed by long flexible polymeric chains dissolved in a Newtonian solvent. It has been well-documented that slow flows of these solutions exhibit purely elastic instabilities, i.e.\ they arise even when the effect of inertia is too small to drive an instability in a Newtonian fluid at the same flow conditions \cite{Larson1992,Shaqfeh1996}. Generally speaking, polymeric flows with curved streamlines exhibit linear instabilities, which are often sub-critical \cite{Groisman2004}. At higher flow rates, dilute polymer solutions exhibit chaotic behaviour,  the so-called \emph{purely elastic turbulence}, which is not related to the usual inertial turbulence \cite{Groisman2000,larson2000}. In flows with straight streamlines, there is no linear instability in the absence of inertia (see \cite{Gorodtsov1967,Wilson1999,Arora2005}, for instance), and the flow exhibits sub-critical transition directly to a chaotic state \cite{Bertola2003,Meulenbroek2004,Morozov2005prl,Morozov2007,Bonn2011,Pan2013} which is, presumably, the same purely elastic turbulence observed at high flow rates in curved geometries.

 Theoretical understanding of the destabilisation mechanism in flows with curved geometries is well-established \cite{Larson1992,Shaqfeh1996,Morozov2007}. It relies on the presence of two ingredients: curved streamlines in the base flow and the velocity gradient across the streamlines. In such situations, polymer molecules stretch in the flow and orient, on average, in the flow direction. The resulting tension in the streamlines, or the so-called \emph{hoop stresses} create extra pressure that increases towards the centre of curvature. At large enough flow rates, this pressure overcomes viscous friction keeping fluid elements on their streamlines and the base flow loses its stability. The resultant flow pattern contains vortices that are typically perpendicular to the direction of the base flow in 3D \cite{Groisman2004}.

Understanding the relationship between the curvature of streamlines and polymeric hoop stresses led Pakdel and McKinley \cite{Pakdel1996,McKinley1996} to formulate the following criterion for the onset of a purely elastic linear instability,
\begin{equation}
\label{eq:PMcK}
\sqrt{\frac{\lambda\,\mathcal{U}}{\mathcal{R}} \frac{N_{1}}{\Sigma_{12}}} \geq M > 0 \;,
\end{equation}
where $\lambda$ is a characteristic relaxation time of elastic stresses in the fluid, $\mathcal{U}$ is a typical velocity of the fluid along a curved streamline, $\mathcal{R}$ is the radius of curvature of that streamline, and $N_1$ and $\Sigma_{12}$ are the first normal stress difference and the shear stress, respectively. A similar instability condition was discussed by Larson, Shaqfeh and Muller \cite{Muller1989,Larson1990,Larson1994} and Shaqfeh \cite{Shaqfeh1996}. Equation \eqref{eq:PMcK} involves two dimensionless groups. First is the ratio between the typical distance travelled along the streamline during one relaxation time, $\lambda\,\mathcal{U}$, and the radius of curvature $\mathcal{R}$. It can be interpreted as a measure of how much the stretched polymers 'feel' the curvature when they follow the streamline. The second group is a ratio between the first normal stress difference $N_1$ and the shear stress $\tau$, and can be viewed as a measure of tension in the streamline. This ratio is sometimes used as an alternative definition of the Weissenberg number $Wi = \lambda \dot\gamma$ \cite{Weissenberg1948}, with $\dot\gamma$ the shear rate (see also below).

This \textit{Pakdel-McKinley criterion} has been experimentally investigated first in a lid driven cavity \cite{Pakdel1996,McKinley1996} where the definition of curvature is less obvious than in a annular geometry. Already in the seminal works on viscoelastic Taylor Couette flow by Muller, Larson and Shaqfeh \cite{Muller1989,Larson1990} the gap width and radii were changed but only over a rather limited range and they found a significant discrepancy to the theoretical predicted scaling. Later on Groisman and Steinberg \cite{Groisman1996, Groisman1998, Groisman1998a} proved the validity of the second part of the Pakdel McKinley criterion, i.e.\ the normal stress to shear stress ratio. However, this can not be fully separated from the first term where the relaxation time and thus the normal stress enters as well. More recently, Zilz \emph{et al.} \cite{Zilz2012} and Poole \emph{et al.} \cite{Poole2013} investigated the visco-elastic instability in a  serpentine micro-channel with a rectangular cross section of different curvatures and confirmed the predicted curvature dependency, while Alves and Poole demonstrated that the Pakdel McKinley criterion can successfully predict the onset of elastic instabilities in smooth contractions of various contraction ratios \cite{Alves2007}.

Ever since the first experimental study by Giesekus \cite{Giesekus1966} that reported the existence of a non-inertial elastic Taylor-Couette instability, numerous studies have been performed, both theoretical \cite{Larson1990, Shaqfeh1992, Avgousti1993, Avgousti1993a, Joo1994,Thomas2006,Thomas2006a,Thomas2009,Khomami2013} and experimental \cite{Baumert1999,Groisman1996,Groisman1998,Groisman1998a,Crumeyrolle2002,Dutcher2013}; see Fardin \emph{et al.} \cite{Fardin2014} for a recent review. Related instabilities were reported in Taylor-Couette flows of worm-like micellar solutions \cite{Becu:2007,Fardin:2009,Fardin:2010,Decruppe:2010,Fielding:2010,Nicolas2012,Mohammadigoushki2017} and in dense colloidal suspensions \cite{Nicolas2016}, although their phenomenology is significantly more complicated due to the rheological properties of the corresponding systems.

Here we present systematic investigation of the geometrical scaling of viscoelastic instabilities in Taylor-Couette geometry by varying the radius of the cylinders and keeping the gap width constant. We will first present our experimental findings and then recall the existing linear stability analysis that are based on the Upper Convected Maxwell (UCM) or Oldroyd-B (O-B) model. These models are not sufficient to describe our fluid rheology and we have to introduce a two-mode model by incorporating the PTT \cite{PhanThien1977} model in our analysis to describe the shear thinning behavior of our fluids.

\section{Experiments}

\subsection{Taylor-Couette setup}
\begin{figure}[h!]
\centering
\includegraphics[trim = 20mm 175mm 10mm 0, clip=true, width=0.9\columnwidth]{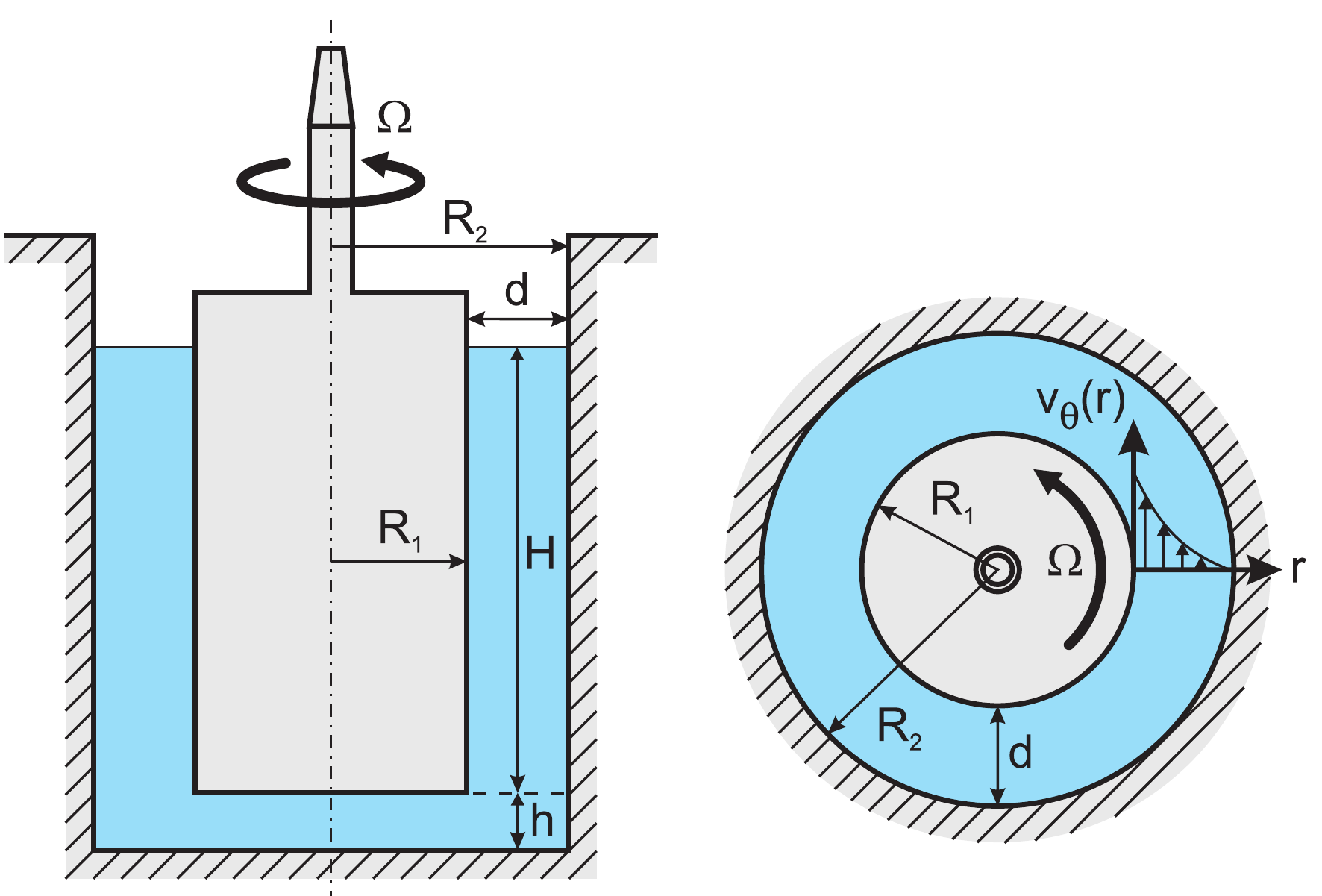}%
\caption{Sketch of the Taylor-Couette cell in side and top view.}%
\label{fig:geometry}%
\end{figure}

Our experiments are performed in the Taylor-Couette geometry, see Figure \ref{fig:geometry} for a sketch of our setup. The radius of the inner cylinder $R_1$ is varied from $2.5$mm to $22$mm and an outer beaker has a radius of $R_2=R_1+d$ (see Tab.\ \ref{tab:geometries}). The gap $d=1$mm is always kept constant giving an explicit change in curvature only. The relative gap width $\varepsilon = d/R_1$ ranges from $0.045$ to $0.4$. The inner (rotating) cylinder is immersed into the fluid down to a distance of $h=10$mm to the bottom end of the outer beaker. The effective contact height of the fluid is $H=73$mm, giving a constant aspect ratio of $\Gamma=H/d=73$. The relative contribution of the fluid disk between the bottom cross-section of the inner cylinder and the bottom of the beaker to the total torque exerted by the fluid on the rotating cylinder does not exceed 1\% of the total torque we measure and can be neglected. We have also checked that in the range of Weissenberg numbers typical for our experiments, there are no instabilities associated with this fluid disk since the onset of either purely elastic or inertial plate-plate instabilities is at much higher shear rates \cite{Larson1992}.

Rotation of the inner cylinder is controlled by a commercial rotational rheometer (MARS II, Thermo Scientific, Karlsruhe, Germany) in controlled rate (CR) mode. The temperature $T$ of the outer Couette cell is kept constant by the use of a closed loop water circuit. The temperature can be stabilized with a precision of $\pm 0.01^{\circ}$C, while the accuracy of the absolute temperature is supposed to be $\pm 0.5^{\circ}$C. The temperature of the inner cylinder was checked to not significantly deviate from the controlled temperature of the beaker, i.e.\ for a setpoint $T=10^{\circ}$C of the outer beaker the temperature of the inner cylinder is $T_1\approx 10.2^{\circ}$C. 

For the flow visualization measurements, a transparent outer beaker has been built and a small amount of anisotropic reflective particles (Kalliroscope) has been added to the solution. As the particles are oriented by following the streamlines of the flow, different flow patterns manifest as regions of different luminance when homogeneously illuminated. The resulting intensity images of the rotating fluid along the whole cylinder axis are captured by a commercial CCD camera.

\begin{table}[th!]
\centering
\begin{tabular}{cccccc}
\toprule
$R_1$ & $R_2$ & \multirow{2}{*}{$\varepsilon = \frac{d}{R_1}$}  \\
  mm & mm & & &\\
\midrule
2.5 & 3.5 & 0.400 \\
3.75 & 4.75 & 0.267 \\
5 & 6 & 0.200 \\
7.5 & 8.5 & 0.133 \\
10 & 11 & 0.100 \\
15 & 16 & 0.067 \\
17.5 & 18.5 & 0.057 \\
20 & 21 & 0.050 \\
22 & 23 & 0.045 \\
\bottomrule
\end{tabular}
\caption{Parameters of the used Taylor-Couette cells. The gap width $d=R_2-R_1=1$mm is the same for all the different setups.}
\label{tab:geometries}
\end{table}

\subsection{Sample preparation and characterisation}
\label{sec:sample_prep}
Highly elastic, long-chained Polyacrylamide (PAAm, molecular weight $5-6$ Mio and $18$ Mio Dalton) molecules dissolved in different Newtonian solvents were used to obtain highly visco-elastic solutions.  Table \ref{tab:polymersolutions} gives an overview of the solutions used in our experiments. The polymer concentration $c_\text{PAAm}$ is varied from $80$ppm to $1200$ppm, and we used either aqueous glycerol or saccharose mixtures of various concentrations as Newtonian solvents. All solutions were prepared according to the following protocol: First, the polymer powder was dissolved in water by moderate shaking and stirring for $24$ hours at ambient temperature. Next, the appropriate amount of glycerol or saccharose was added and the whole solution was gently stirred for another period of $24$ hours. While the glycerol solutions are very robust against chemical degradation, the sugar solutions get rapidly infested by mold, as they are a perfect culture medium for bacteria. Therefore, all the measurements with sugar-based solvents were performed immediately after finishing the preparation.
\begin{table}[!th]
\centering
\begin{tabular}{cccccc}
\toprule
\multirow{2}{*}{name} & $c_\text{PAAm}$ & $M_w$ & \multicolumn{2}{c}{solv.\ (X+H$_2$O)} & $T$\\
 & (ppm) & $(10^{6}$ Da) & X & \% & ($^{\circ}$C)\\
\midrule
P150$_\text{G80}$ & 150 & \multirow{4}{*}{5--6* } & glyc. & 80 & 10 \\
P600$_\text{G80}$ & 600 &  & glyc. & 80 & 10 \\
P1200$_\text{G80}$ & 1200 &  & glyc. & 80 & 10 \\
P500$_\text{S58}$ & 500 &  & sacch. & 58 & 10\\
\hline
P80$_\text{S64}$ & 80 & \multirow{2}{*}{18**}  & sacch. & 64 & 22 \\
P150$_\text{S65.6}$ & 150 & & sacch. & 65.5 & 23 \\
\bottomrule
\end{tabular}
\caption{PAAm solutions used in this study (Sigma-Aldrich No.\ 92560 (*), Polysciences No.\ 18522 (**)).  Solutions with the lower molecular weight were studied at a lower temperature than the high-molecular weight samples in order to increase the stress signal and to make the instability more pronounced.}
\label{tab:polymersolutions}
\end{table}

Rheological characterisation of the solutions was performed in the cone-plate ($C60/2^\circ$) geometry with the same rheometer as the Taylor-Couette measurements. Figure \ref{fig:PAAm600ppm_characterization} shows representative steady-shear data of the apparent shear viscosity $\eta(\dot\gamma)$ and the first normal stress difference $N_1(\dot\gamma)$ for the P600$_\text{G80}$ solution. All our solutions exhibit a moderate degree of shear thinning, and the first normal stress difference $N_1(\dot\gamma)$ is approximately proportional to the square of the shear rate.

\subsection{Fitting the rheology: a hybrid model}
\label{subject:rheology}
As a first approximation, our solutions can be described by the Oldroyd-B constitutive equation that expresses the total stress in the fluid, $\bm{\Sigma}$, as a sum of an isotropic pressure $p$, a Newtonian contribution with the viscosity $\eta_s$, and a polymeric contribution $\bm{\tau}_u$ that obeys the Upper-Convected Maxwell (UCM) model \cite{Larson1992,Bird1987}
\begin{gather}
\label{eq:oldroyd-b}
\bm{\tau}_u + \lambda_u \ucd{\bm{\tau}_u} = \eta_u \left(\nabla\vec{v} + \nabla\vec v^\dagger\right).
\end{gather}
Here, $\vec{v}$ is the velocity of the fluid, $\lambda_u$ and $\eta_u$ are the Maxwell relaxation time and the polymeric viscosity of the UCM model, respectively; $^\dagger$ denotes transpose of a matrix. The upper-convected derivative of a second-rank tensor is given by \cite{Bird1987,Bird1995}
\begin{gather}
\ucd{\bm{\tau}} = \frac{\partial \bm{\tau}}{\partial t} + \vec{v}\cdot\nabla \bm{\tau} - \nabla\vec v^\dagger\cdot \bm{\tau} - \bm{\tau}\cdot \nabla\vec v.
\end{gather}
Additionally, the fluid satisfies the momentum-balance and incompressibility equations
\begin{gather}
\label{eq:ns}
\rho \left( \frac{\partial\vec{v}}{\partial t} + \vec{v} \cdot \nabla \vec{v} \right)
 =\nabla \cdot \bm{\Sigma}, \\
\label{eq:incomp}
\nabla \cdot \vec v = 0.
\end{gather}

\begin{figure}[!th]
\includegraphics[width=\columnwidth]{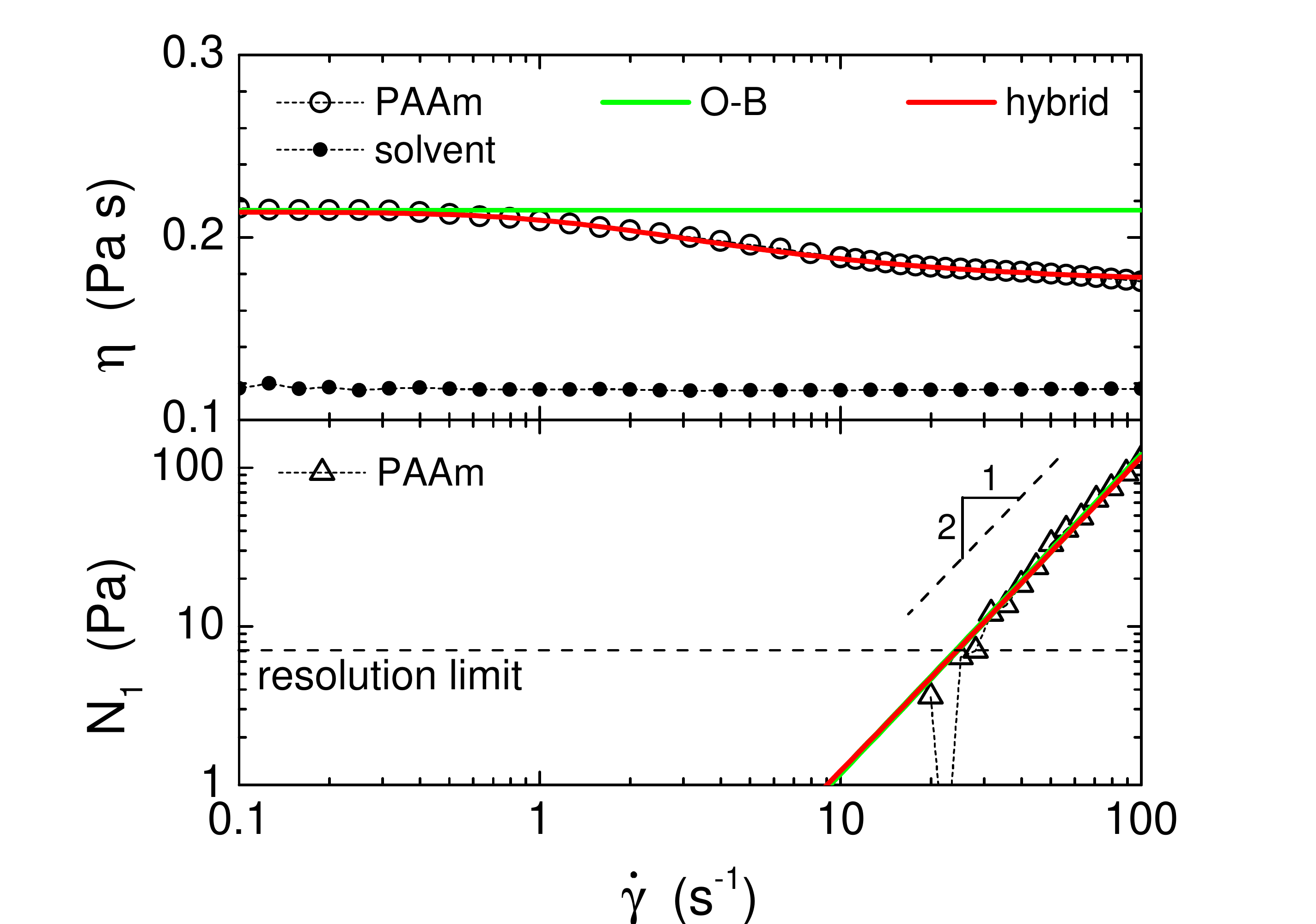}%
\caption{Shear rate sweep measurement of the P600$_\text{G80}$ solution and its respective solvent in terms of the viscosity $\eta(\dot\gamma)$ (\textit{top}) and first normal stress difference $N_1(\dot\gamma)$ (\textit{bottom}). The data are fitted according to the Oldroyd-B (\textit{green}) and the hybrid model (\textit{red}).}%
\label{fig:PAAm600ppm_characterization}%
\end{figure}

According to Eq.~\eqref{eq:oldroyd-b}, 
an Oldroyd-B fluid in steady simple shear flow has a constant total shear viscosity $\eta_s + \eta_u$, and the first normal stress difference $N_1=2 \lambda_u \eta_u \dot\gamma^2$ \cite{Bird1987}. While the latter expression correctly captures the quadratic scaling of $N_1$ with the shear rate observed in our rheological measurements (see Fig.~\ref{fig:PAAm600ppm_characterization}, for example), the shear-rate-independent total viscosity of the Oldroyd-B model is inconsistent with the observation of moderate shear-thinning for all our solutions. Since our goal is to quantitatively assess how predictions of the Pakdel-McKinley condition and the linear stability analysis compare with the experimentally measured onset of purely elastic instabilities, we need a constitutive equation that accurately describes the rheology of our solutions.

In order to compensate for the shortcomings of the Oldroyd-B equation, we use a multi-mode approach \cite{Bird1987} and model the total stress in the fluid $\bm\Sigma$ as 
\begin{gather}
\label{eq:total_stress}
\bm{\Sigma} = -p\,\bm{\delta} + \eta_s  \left(\nabla\vec{v} + \nabla\vec v^\dagger\right) + \bm{\tau}_u + \bm{\tau}_p,
\end{gather}
where $\bm{\delta}$ is a second-rank identity tensor.
As before, the second and the third terms are the Newtonian and UCM stress tensors, correspondingly. The last term, $\bm{\tau}_p$, obeys the simplified, linearised formulation (sPTT \cite{Mirzazadeh2005}) of the PTT model \cite{PhanThien1977}
\begin{gather}
\label{eq:sPTT}
\bm{\tau}_p \left(1 + \alpha\frac{\lambda_p}{\eta_p} \text{tr}\bigl(\bm{\tau}_p\bigr)\right) + \lambda_p\ucd{\tau_p} = \eta_p \left(\nabla\vec v + \nabla\vec v ^\dagger \right).
\end{gather}
Similar to the UCM model, an sPTT fluid is characterised by a relaxation time $\lambda_p$ and viscosity $\eta_p$; additionally, Eq.~\eqref{eq:sPTT} contains a parameter $\alpha\ge0$ that controls the degree of shear-thinning exhibited by this fluid. 

In steady simple shear flow, the only non-zero components of the sPTT stress $\bm{\tau}_p$ are the shear stress  $\tau_{p,12}$ and the normal stress $\tau_{p,11}$ given by
\begin{gather}
\label{eq:PTTeta}
\tau_{p,12}(\dot\gamma) \left(1 + 2\alpha\left(\frac{\lambda_p}{\eta_p}\right)^2 {\tau_{p,12}}(\dot\gamma)^2\right) = \eta_p \dot\gamma \;,\\
\tau_{p,11}(\dot\gamma) = 2 \frac{\lambda}{\eta_p} {\tau_{p,12}(\dot\gamma)}^2.
\end{gather}
As can be seen from these equations, at low stresses, the sPTT model reproduces the Oldroyd-B rheology, while for large stresses, one obtains $\tau_{p,12}\sim\dot\gamma^{1/3}$ and $\tau_{p,11}\sim\dot\gamma^{2/3}$, implying significant shear-thinning of both viscosity and the first-normal stress difference. 

Within our two-mode \emph{hybrid model}, the steady shear rheology is then described by
\begin{align}
\eta(\dot\gamma) &= \eta_s + \eta_u + \frac{\tau_{p,12}(\dot\gamma)}{\dot\gamma},\\
N_1(\dot\gamma) &= 2 \lambda_u \eta_u \dot\gamma^2 + \tau_{p,11}(\dot\gamma).
\end{align}
Our modelling strategy is based on ensuring that the normal-stress difference is dominated by the Oldroyd-B component, while the total viscosity is dominated by the sPTT viscosity and, hence, exhibits shear-thinning. An example of the fit of our hybrid model to the data for the P600$_\text{G80}$ solution is shown in Fig.~\ref{fig:PAAm600ppm_characterization}, demonstrating a fairly good agreement.
\begin{table*}[!tbh]
\centering
\scriptsize
\begin{tabular}{ccccccccccc}
\toprule
\multirow{2}{*}{solution} & $\eta_s$ & $\eta_u$ & $\eta_p$ & $\lambda_u$ & $\lambda_p$ & $\alpha{\left(\lambda_p\right)}^2$ & \multirow{2}{*}{$n$} & $\lambda^\text{rheo}$ & \multirow{2}{*}{$\beta$} \\
 & (mPa\,s) & (mPa\,s) &  (mPa\,s) & (ms) & (ms) & (s$^2$) &  & (ms) \\
\midrule
P150$_\text{G80}$   & $119 \pm 2$ & $11.8 \pm 0.1$  & $7.6 \pm 0.2$  & $51 \pm 3$ & $ (0.9 \pm 1.5)\cdot 10^3$ & $ 0.028 \pm 0.013$ & $0.91 \pm 0.06$ & $4.8 \pm 0.4$ & $0.86 \pm 0.03$  \\
P600$_\text{G80}$   & $119 \pm 2$ & $57 \pm 1$  & $40 \pm 1$ & $99 \pm 5$ & $23 \pm 39$ & $0.066 \pm 0.014$ & $0.96 \pm 0.03$ & $33 \pm 2$ & $0.55 \pm 0.02$ \\
P1200$_\text{G80}$  & $119 \pm 2$ & $123 \pm 1$ & $114 \pm 1$ & $134 \pm 8$ & $229 \pm 323$ & $0.094 \pm 0.002$ & $0.97 \pm 0.03$ & $68 \pm 4$ & $0.33 \pm 0.01$  \\
P500$_\text{S58}$   & $83.1 \pm 0.2$ & $38 \pm 3$ & $28 \pm 1$ & $67 \pm 6$ & $126 \pm 204$ & $0.043 \pm 0.006$ & $0.91 \pm 0.02$ & $22 \pm 1$ & $0.56 \pm 0.02$ \\
\midrule
P80$_\text{S64}$  & $140 \pm 2$ & $15 \pm 3$ & $22 \pm 1$ & $242 \pm 49$ & $ (3.1 \pm 4.8) \cdot 10^3$ & $0.68 \pm 0.20$ & $0.56 \pm 0.06$ & $64 \pm 15$ & $0.79 \pm 0.04$ \\
P150$_\text{S65.5}$ & $105.6 \pm 0.5$ & $50 \pm 5$ & $69 \pm 4$  & $447 \pm 46$ & $(4.2 \pm 7.4)\cdot 10^3$ & $1.30 \pm 0.47$ & $0.94 \pm 0.06$ & $153 \pm 11$ & $0.47 \pm 0.02$   \\
\bottomrule
\end{tabular}
\caption{Rheological parameters of the different PAAm solutions based on the two-mode hybrid model.}
\label{tab:solpara}
\end{table*}

In Table~\ref{tab:solpara}, we present the model parameters, $(\eta_s, \eta_u, \lambda_u, \eta_p, \lambda_p, \alpha)$, obtained by fitting our hybrid model to the steady-state shear rheology of the solutions listed in Table~\ref{tab:polymersolutions}. In general, most of the parameters of the hybrid model show systematic variation with the polymer concentration and the solvent viscosity, the only exception being the sPTT relaxation time $\lambda_p$. 
The reason for wide variations of $\lambda_p$ and the associated large standard deviation for this parameter is the fact that the sPTT component of the model predominantly contributes to the shear stress (viscosity) and not to the normal stresses. Shear-thinning viscosity of our solutions allows us to determine the combination $\alpha \lambda_p^2$ that also shows systematic variation with the polymer concentration, see Table~\ref{tab:solpara}. In order to disentangle $\alpha$ and $\lambda_p$ in this combination, one needs to use the normal-stress data. However, since $N_1$ is well-described by the UCM component of the model, the absolute value of $\lambda_p$ is difficult to determine resulting in large errors in its values in Table~\ref{tab:solpara}. Nevertheless, this does not significantly affect the rheology of the model since the sPTT component is used as, essentially, a power-law fluid with almost no normal stresses.

\begin{figure}[!th]
\includegraphics[width=\columnwidth]{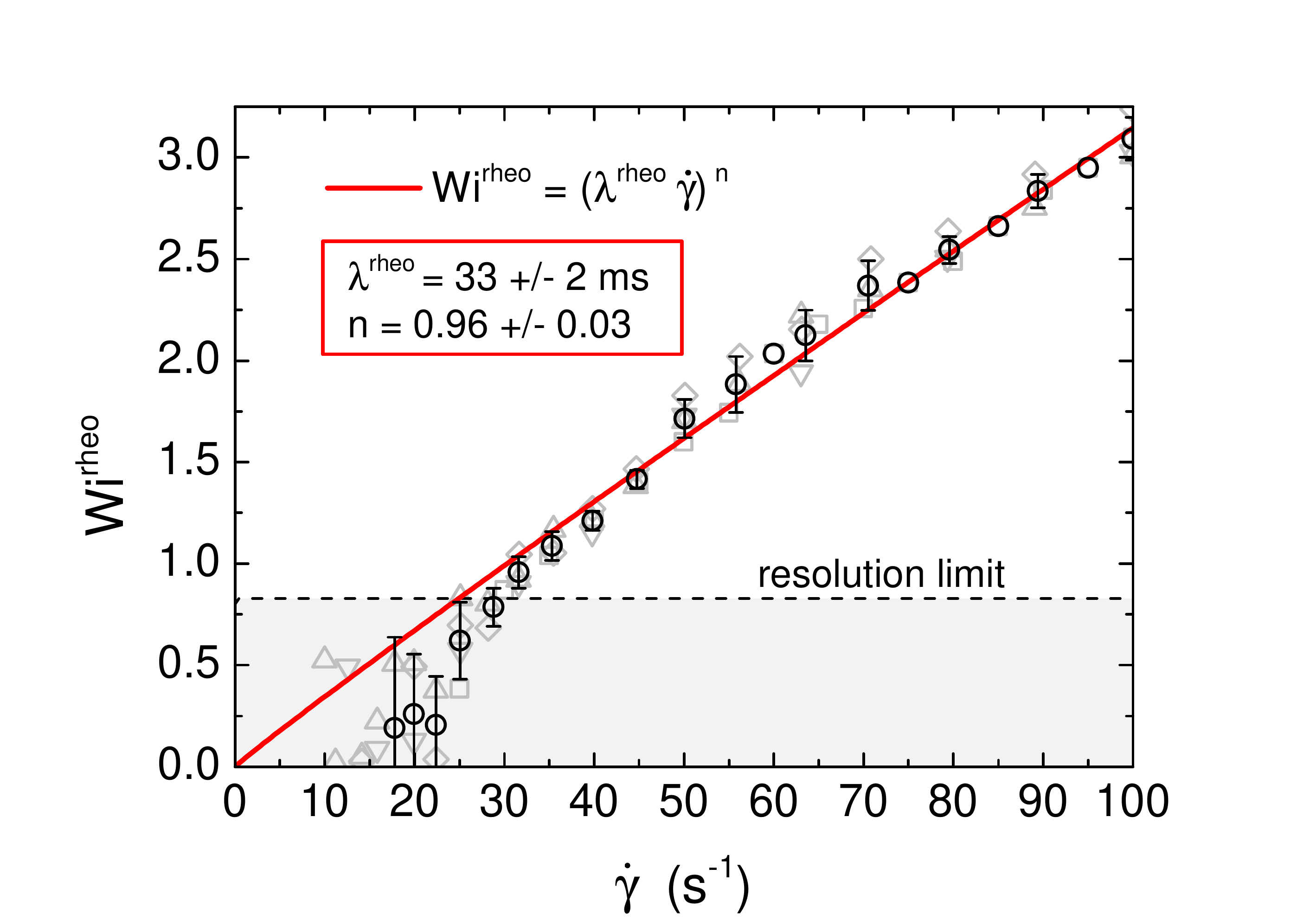}
\caption{Rheological Weissenberg number as a function of the applied shear rate for the P600$_\text{G80}$ solution. Different symbols refer to independent sets of measurements in the cone-and-plate setup and circles give the averaged data.}
\label{fig:WirheoPAAm600G80}
\end{figure}

In what follows, we will be comparing the results of our experiments with the predictions of the Pakdel-McKinley criterion and the linear stability analysis. This comparison is only meaningful if we can define the Weissenberg number in our theory and experiments in the same way. Formally, it can be defined based on the UCM-component relaxation time $\lambda_u$ extracted from the data with the help of our hybrid model since that component is responsible for reproducing the normal-stress behaviour. However, we feel that such a definition would be too model-dependent and, additionally, ignore the shear-thinning nature of our fluid. Another approach is motivated by the observation that for the UCM model, the Weissenberg number can be written as $N_1/2\tau_{u,12}$. For a shear-thinning fluid, a popular definition of the Weissenberg number is to use the UCM expression at a particular shear rate, i.e. $N_1(\dot\gamma)/2\tau_{u,12}(\dot\gamma)$. This approach relies on one's ability to separate the polymer shear stress from the total shear stress in the fluid, and, naively, the difference between the two can be approximated by $\eta' \dot\gamma$, where $\eta'$ is the viscosity of the pure Newtonian solvent in the absence of polymers. This is, however, only an estimate since the presence of polymers certainly changes the viscosity of the fluid even if the polymers are not significantly stretched \cite{Zell2010}, and, in general, $\eta_s$ from Eq.~\eqref{eq:total_stress} is larger than $\eta'$. To avoid this complication, we interpret our experimental measurements in terms of a \emph{rheological Weissenberg number} based on the values of the normal stresses and the \emph{total} shear stress at a given shear rate:
\begin{equation}
\label{eq:Wi_rheo}
Wi^\text{rheo}(\dot\gamma) = \frac{N_{1}(\dot\gamma)}{2|\Sigma_{12}(\dot\gamma)|}.
\end{equation}
In Figure \ref{fig:WirheoPAAm600G80}, we present the rheological Weissenberg number for the P600$_\text{G80}$ solution as calculated from the rheology data, revealing a power law dependence on the shear rate in the relevant region between $30$s$^{-1}$ and $100$s$^{-1}$, i.e.
\begin{equation}
\label{eq:Wirheo}
Wi^\text{rheo} = \left(\lambda^\text{rheo}  \dot\gamma\right)^n = (33 \pm 2\text{ms} \cdot \dot\gamma)^{0.96 \pm 0.03} ~.
\end{equation}
with a power-law exponent $n$ and an alternative (relaxation) time $\lambda^\text{rheo}$. The deviation from the linear dependency on the shear rate is rather weak for this solution since the shear thinning of the viscosity as well as the deviations of the normal stress from the quadratic scaling with the shear rate are moderate. This is true for most of the investigated solutions (cf.\ Table~\ref{tab:solpara}).

The results of our theory will also be reported in terms of the rheological Weissenberg number defined with the help of the hybrid model as 
\begin{equation}
\label{eq:Wi_rheo_hybrid}
Wi^\text{rheo}(\dot\gamma) = \frac{2 \lambda_u \eta_u \dot\gamma^2 + \tau_{p,11}(\dot\gamma)}{2\left|\left(\eta_s + \eta_u\right)\dot\gamma + \tau_{p,12}(\dot\gamma)\right|}.
\end{equation}

\subsection{Taylor-Couette measurements}

All our solutions were investigated with respect to their transition to elastic instability in the various Taylor-Couette cells of different relative gap widths $\varepsilon$. For each Taylor-Couette cell we performed shear rate sweep measurements, that is, the shear rate was increased stepwise, starting from $\dot\gamma=1$s$^{-1}$ up to $\dot\gamma=200$s$^{-1}$ in steps of $1$s$^{-1}$.
\begin{figure}[!th]
\includegraphics[trim = 10mm 5mm 30mm 20mm, clip=true, width=\columnwidth]{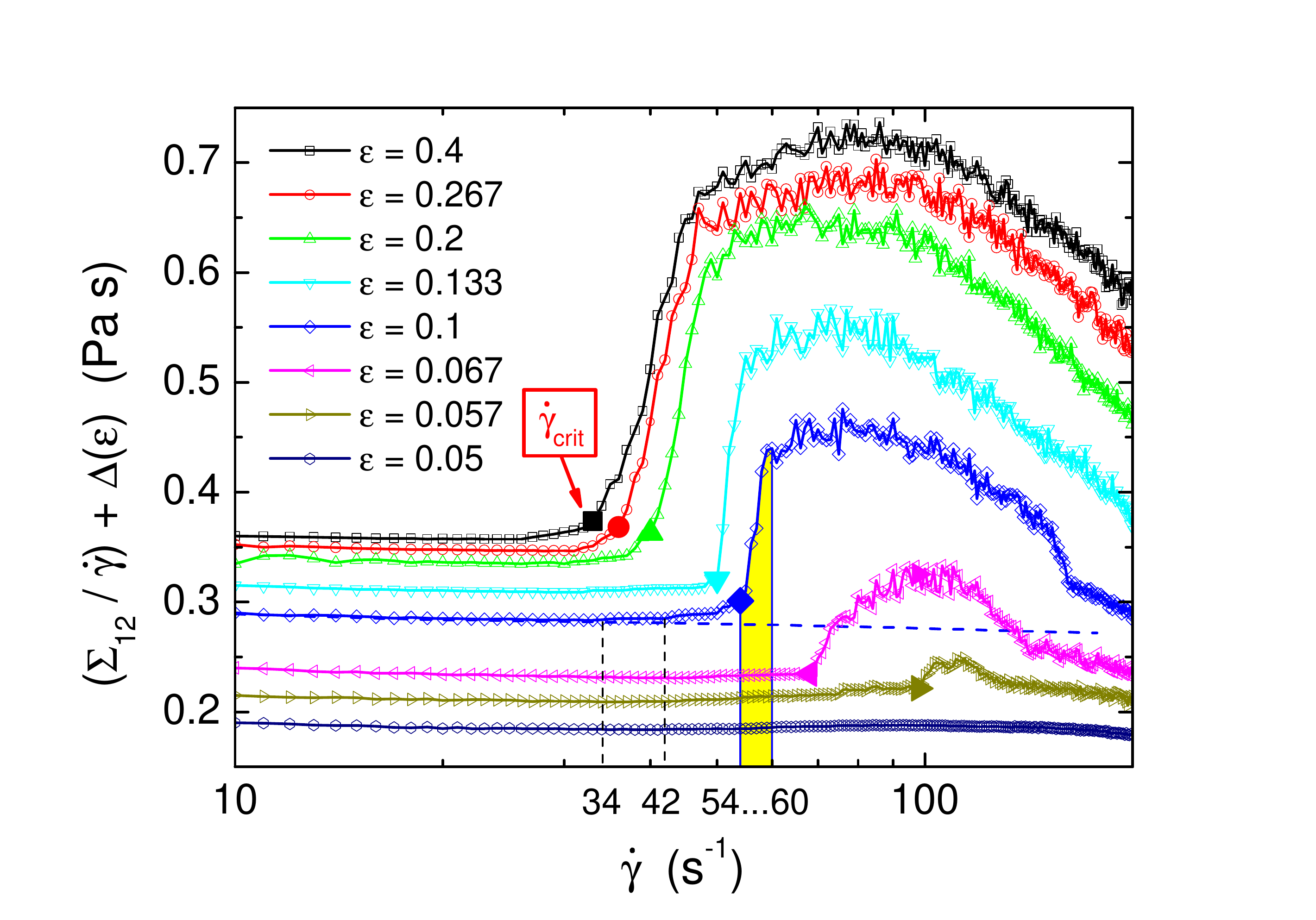}%
\caption{Taylor-Couette data of the P600$_\text{G80}$ solution. The curves are vertically shifted in proportion to the respective radius of the inner cylinder relative to the biggest one in use, i.e.\ $\Delta(\varepsilon) = (20 - \varepsilon^{-1}) \cdot 0.01$Pa\;s. The blue dashed line indicates the viscometric flow curve. For $\varepsilon = 0.1$, the flow has been additionally visualized as presented in Fig.\ \ref{fig:spacetime} (yellow area).   }%
\label{fig:TC_P600}%
\end{figure}
Each shear rate was kept constant for $\Delta t = 10$s and the ratio of the measured shear stress and shear rate $\Sigma_{12}(\dot\gamma)/ \dot\gamma $ was averaged over the last $5$s of each time period $\Delta t$. Figure \ref{fig:TC_P600} shows representative flow curves (vertically shifted) of the P600$_\text{G80}$ solution measured at $T=10^{\circ}$C.  All curves show a common behaviour: for low shear rates, corresponding to the case of viscometric, purely azimuthal Couette flow, the extracted quantity gives a slightly shear thinning viscosity of the solution as measured in the cone-and-plate setup. When increasing the shear rate above some ill defined value, the signal starts to slightly increase above the viscometric flow curve. Subsequently, the curve sharply increases at a critical shear rate $\dot\gamma_\text{crit}$. We refer to this point as the onset of elastic instability in agreement with a visualization of non-trivial flow patterns, which suddenly spread across the whole cylinder at $\dot\gamma_\text{crit}$ (cf.\ red box in space-time plot in Fig.\ \ref{fig:spacetime}). The value of $\dot\gamma_\text{crit}$ systematically depends on the relative gap width $\varepsilon$ as well as on the fluid parameters. Figure \ref{fig:summary} summarizes all the transition points in terms of the critical modified Weissenberg number $\sqrt{\varepsilon} Wi^\text{rheo}_\text{crit}$ as calculated via Eq.\ \eqref{eq:Wirheo}. The error bars give a worst-case error estimate, accounting for errors in the normal stress and viscosity data as well as in the localization of the critical shear rate. When the annular gap is completely occupied by secondary flow patterns, the sharp incline of the stress data in Fig.\ \ref{fig:TC_P600} stops and the curves slowly rise to some maximum before decreasing again. This is due to the degradation of the polymer solutions, which occurs under the influence of strong deformation due to strong secondary flow \cite{Groisman2000,Groisman2001,Groisman2004,Elbing2009,Owolabi2017}. We also note here that the contribution of the instability to the torque on the inner cylinder, i.e. the rise of the effective viscosity above its viscometric value in Fig.~\ref{fig:summary}, is larger for larger relative gap width $\epsilon$.
\begin{figure}[!th]
\centering
\includegraphics[trim = 0 0 0 0 , clip=true, width=0.93\columnwidth]{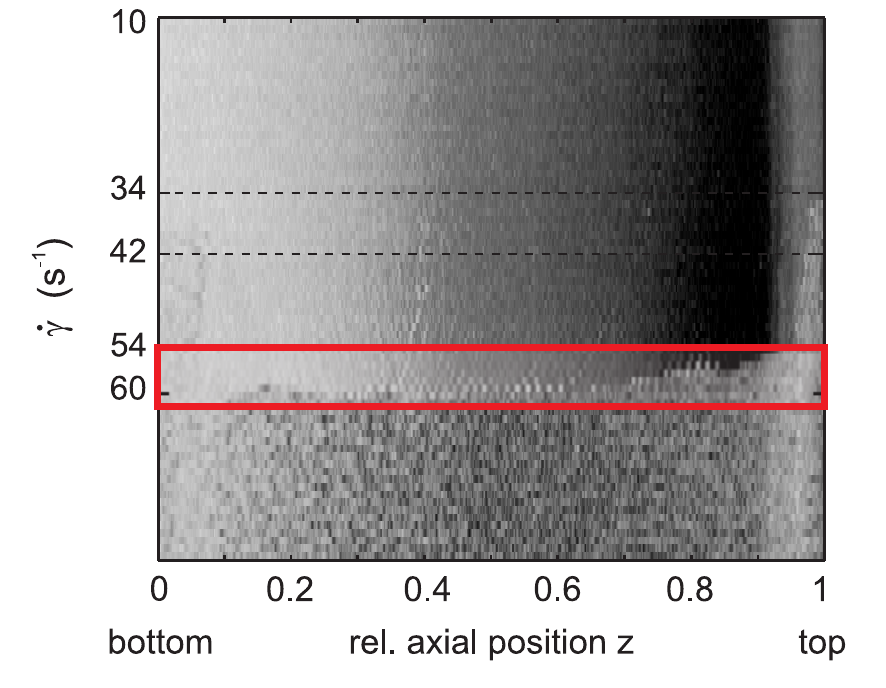}
\caption{Visualization of the fluid flow (face view) during a standard shear rate sweep measurement of P600$_\text{G80}$ for $\varepsilon = 0.1$. Between $\dot\gamma_\text{crit} = 54$s$^{-1} \leq \dot\gamma \leq 60$s$^{-1}$ (red box), non-trivial flow patterns spread across the whole cylinder axis.}
\label{fig:spacetime}
\end{figure}
\begin{figure}[!th]
\centering
\includegraphics[trim = 0mm 0mm 0mm 20mm, clip=true, width=\columnwidth]{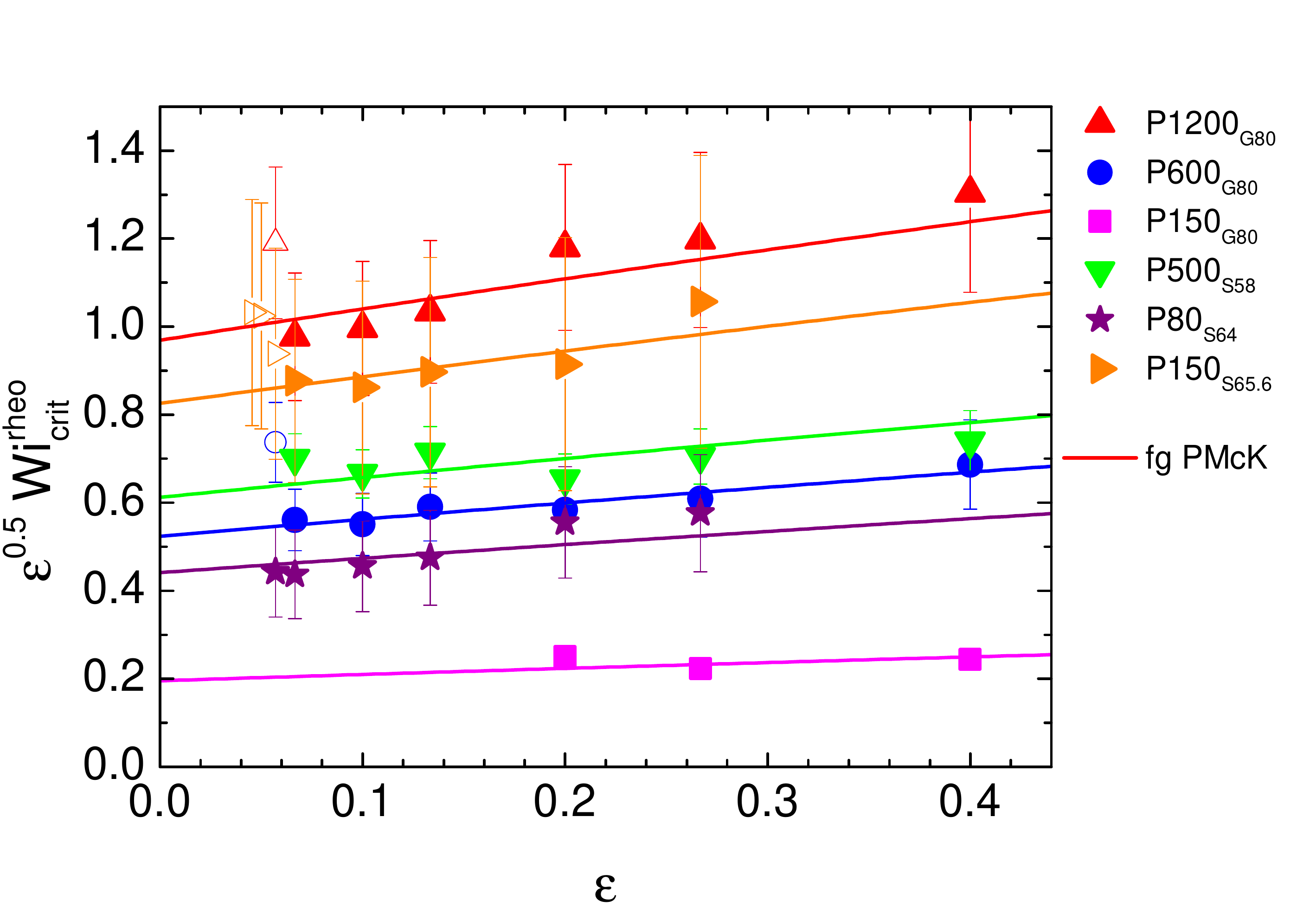}%
\caption{Summary of the critical modified rheological Weissenberg numbers mapping the onset of elastic instability in the Taylor-Couette setups of different gap widths $\varepsilon$. Open symbols denote data points at small $\varepsilon$ where a clear determination of the onset became difficult. Those data where not taken for the fit. The geometrical scaling is well described by the Pakdel-McKinley criterion \eqref{eq:PMcK_fg_rheo} for moderate to high values of $\varepsilon \gtrsim 0.05$.}%
\label{fig:summary}%
\end{figure}

In the following sections we compare our measurements against the predictions of the Pakdel-McKinley criterion and the linear stability analysis of the hybrid model, and demonstrate that while the Pakdel-McKinley condition successfully describes variation of the onset of the elastic instability with $\epsilon$, surprisingly, the linear stability analysis fails, both qualitatively and quantitatively. 

\section{Pakdel-McKinley criterion for Taylor-Couette flow}
\label{seq:PM}

As mentioned in the Introduction, Pakdel and McKinley have developed a condition for the onset of a linear instability in shear flows with curved streamlines \cite{Pakdel1996,McKinley1996}, and the most general form of that condition is summarised in Eq.~\eqref{eq:PMcK} \cite{Morozov2007}. In the small-gap approximation, $\varepsilon \ll 1$, the base Taylor-Couette flow is approximated by a linear shear with a constant shear rate across the gap given by $\dot\gamma_{sg}=\Omega R_1/d$, where the subscript refers to the small-gap approximation. First, we neglect the effect of shear-thinning of our solutions, as was done by Pakdel and McKinley \cite{Pakdel1996,McKinley1996}, and base the following discussion on the Oldroyd-B model; it will be incorporated in our instability condition later on. Using the Oldroyd-B expressions for the stresses in linear shear in our definition of the rheological Weissenberg number, Eq.~\eqref{eq:Wi_rheo}, we obtain $Wi^\text{rheo}=\left(1-\beta\right)\lambda_u \dot\gamma_{sg}$, where $\beta=\eta_s/\eta$ is the ratio of the solvent and total viscosities of the solution. Choosing $\mathcal{U} = \Omega R_1$ for the characteristic velocity along a streamline and the radius of curvature $\mathcal{R} = R_1$, the instability condition \eqref{eq:PMcK} is given by
\begin{equation}
\label{eq:PMcK_TC_OB}
\sqrt{\varepsilon} Wi^\text{rheo} \geq \sqrt{\frac{1-\beta}{2}}M,
\end{equation}
or in terms of the usual Weissenberg number, $Wi=\lambda_u \dot\gamma_{sg}$, 
\eqref{eq:PMcK} is given by
\begin{equation}
\label{eq:PMcK_TC_OB_normalWi}
\sqrt{\varepsilon} Wi\geq \frac{M}{\sqrt{2(1-\beta)}}.
\end{equation}

The combination $\sqrt{\varepsilon} Wi^\text{rheo}$ is often referred to as the \textit{modified Weissenberg number} \cite{Larson1992}. For the simplest case of a UCM fluid ($\beta=0$), Eq.~\eqref{eq:PMcK_TC_OB} simplifies to $2\varepsilon (Wi^\text{rheo})^2 \geq {M}^2$, that resembles the criterion for the linear instability in the Newtonian case with the Reynolds number replaced by the Weissenberg number, and the left hand side being the Taylor number $Ta$ \cite{Larson1992}.

Eq.~\eqref{eq:PMcK_TC_OB} defines the onset of an elastic linear instability in terms of a critical modified Weissenberg number as a function of the relative solvent viscosity and a constant $M$, assumed to be universal for a given type of flow. In the small-gap approximation of a Taylor-Couette flow of an Oldroyd-B fluid, the modified Weissenberg number does not explicitly depend on the geometrical parameter $\varepsilon$.

While providing a simple instability criterion in the small-gap approximation, Eq.~\eqref{eq:PMcK_TC_OB} is not applicable to our experiments due to the dimensions of our Taylor-Couette cells.
Indeed, even at the smallest gap $\varepsilon = 0.045$ used in our experiments, the shear rates varies about $10\%$ across the gap, while for $\varepsilon = 0.4$, the variation is around $40\%$.
This indicates that for the Taylor-Couette geometries used in our experiments (and the previous ones presented in the literature) the finite width of the gap should be considered.

The adaption of the Pakdel-McKinley criterion to the general case of finite gap width is straightforward. In the Oldroyd-B model, the velocity profile in the gap, $v_\theta(r)$, is given by its Newtonian expression, $v_\theta(r)=A\,r + B/r$, where constants $A$ and $B$ are determined from the boundary conditions $v_\theta(R_1) = \Omega R_1$ and $v_\theta(R_1+d) = 0$. The shear rate in the gap is then $\partial v_\theta(r)/\partial r - v_\theta(r)/r$, and is maximum at the inner cylinder $r=R_1$. Rewriting Eq.~\eqref{eq:PMcK_TC_OB} in terms of this maximum shear rate yields
\begin{equation}
\label{eq:PMcK_fg_rheo}
\sqrt{\varepsilon}Wi^\text{rheo}(\varepsilon;\beta) \geq f(\beta) \sqrt{1 + \frac32\varepsilon + \frac{\varepsilon^2}{2(\varepsilon+2)} },
\end{equation}
where
\begin{equation}
\label{eq:fbeta}
f(\beta)=\frac{M}{\sqrt{2}}\sqrt{1-\beta}\;.
\end{equation}
\begin{figure}[!th]
\includegraphics[trim = 10mm 0mm 10mm 10mm, clip=true, width=1\columnwidth]{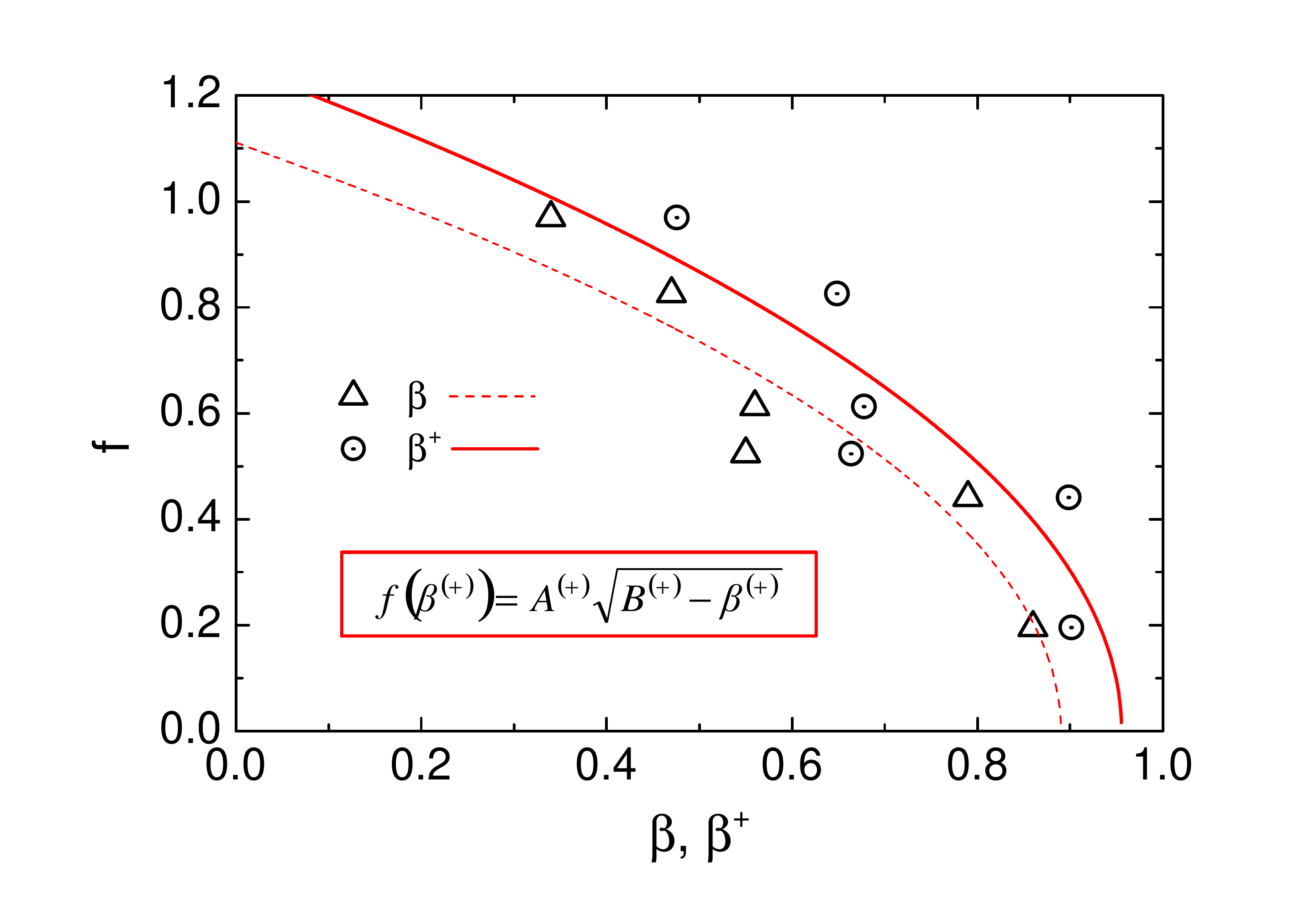}%
\caption{The critical parameter $f$ from the Pakdel-McKinley criterion \eqref{eq:PMcK_fg_rheo} as a function of the zero shear rate viscosity ratio $\beta$ (triangles and dashed line) and the shear thinning viscosity ratio $\beta^+ = \eta_s/\eta^+ = \eta_s/\eta(\dot\gamma)$ (circles and line). Symbols refere to experimental data points, the lines to the fitting function $f\left(\beta^{(+)}\right)=A^{(+)}\sqrt{B^{(+)}-\beta^{(+)}}$.}%
\label{fig:M_vs_beta}%
\end{figure}
In Fig.~\ref{fig:summary} we compare the prediction of Eq.~\eqref{eq:PMcK_fg_rheo} with the experimental data. Since the Pakdel-McKinley condition contains an unknown constant, we use the scaling factor $f(\beta)$ in Eq.~\eqref{eq:PMcK_fg_rheo} as a fitting parameter. As can be seen from Fig.~\ref{fig:summary}, the Pakdel-McKinley criterion reproduces the experimental results fairly well for 
$\varepsilon \gtrsim 0.05$. 

Now we turn our attention to the dependence of the fitting parameter $f$ on $\beta$ and compare it with the actual prediction of the Pakdel-McKinley condition, Eq.~\eqref{eq:fbeta}. In Fig.~\ref{fig:M_vs_beta} we plot the values of $f$ obtained from fitting the experimental onset data to Eq.~\eqref{eq:PMcK_fg_rheo} for each solution as a function of $\beta = \eta_s / \eta(0)$. Here $\eta_s$ is the nominal solvent contribution to the total viscosity and $\eta(0)$ is the zero-shear rate viscosity of each solution, see Table~\ref{tab:solpara}. The data can be approximated by a function $f(\beta) = A \sqrt{B-\beta}$ with parameters $A = 1.2 \pm 0.1$ and $B = 0.89 \pm 0.03$, which differs from the prediction Eq.~\eqref{eq:fbeta}. 

A better agreement can be obtained by accounting explicitly for the shear-thinning properties of the fluids. As discussed in Section~\ref{subject:rheology}, viscosity of our solutions is moderately shear-thinning, while the first normal-stress difference is well-described by the Oldroyd-B model. Therefore, in Fig.~\ref{fig:M_vs_beta} we also plot the fitted values of $f$ against 
the shear-thinning viscosity ratio $\beta^+ = \eta_s/\eta^+ = \eta_s/\eta(\dot\gamma)$, where $\eta(\dot\gamma)$ is the actual viscosity of the solution at the shear rate where an instability was detected; a similar procedure was used by Casanellas \emph{et al.} \cite{Casanellas2016}. Fitting these data to the function $f(\beta^+) = A^+ \sqrt{B^+ - \beta^+}$ gives $A^+ = 1.3 \pm 0.2$ and $B^+ = 0.96 \pm 0.05$. Within the experimental error, this function is identical to the prediction of the Pakdel-McKinley expression, Eq.~\eqref{eq:fbeta}. 

From our data we conclude that the universal constant $M$ in the Pakdel-McKinley criterion is equal to
\begin{equation}
M = 1.8 \pm 0.3.
\label{Meqtmp}
\end{equation}
In terms of the overall rheological scaling, our results are in agreement with experiments of Groisman and Steinberg \cite{Groisman1998}, who measured the onset of disordered oscillations (DO) at a fixed value of $\varepsilon = 0.255$ for different polymer solution at various temperatures. However, they extracted a higher value of $M = 3.58$ that can probably be attributed to the fact that they used a different relaxation time that was extracted from linear oscillatory measurements.

\section{Linear stability analysis}
\label{seq:LSA}

In this section we assume that the threshold we observed experimentally corresponds to the loss of stability by the base, laminar flow with respect to infinitesimal perturbations, \emph{i.e.} we observed a linear instability. As usual, occurrence of a linear instability is identified theoretically by introducing a small perturbation to the base state in the equations of motion, which in the case of Taylor-Couette flow takes the following form: $\left\{\bm{\delta v}(r),\delta p(r),\bm{\delta\Sigma}(r)\right\}e^{i k z}e^{i m \theta}e^{\mu t}$. Here, $k$ and $m$ are the axial and azimuthal wavenumbers, correspondingly, and $\mu$ is the eigenvalue that, in general, is a function of the Reynolds number $Re$, Weissenberg number $Wi$, and $k$ and $m$. For given $m$ and $k$, the threshold of a linear instability is then determined by the point where the real part of $\mu$ becomes positive. 

Our goal here is to compare the experimental threshold to predictions of the linear stability analysis and assess whether it agrees with the Pakdel-McKinley criterion. First we summarise existing results on the linear stability of Taylor-Couette flow. Most of the previous experimental work addressing the scaling of the instability threshold with the parameters of the Taylor-Couette geometry was performed with Boger fluids, and the corresponding linear stability analysis is for the Oldroyd-B model. Then we present the results of the linear stability analysis for our shear-thinning hybrid model and compare it with the predictions of the Pakdel-McKinley criterion.

\subsection{Review of previous studies}

The first systematic linear stability analysis of Taylor-Couette flow was performed by Larson, Shaqfeh and Muller  \cite{Larson1990}. Employing the small-gap (sg) approximation they derived an analytic expression for the instability threshold with respect to axisymmetric modes ($m=0$):
\begin{equation}
\label{eq:PMcK_sg_LSA}
\textit{small gap:} \qquad \sqrt{\varepsilon}Wi \geq \frac{5.92 \pm 0.02}{\sqrt{K(\beta)}}.
\end{equation}
Here,
\begin{align}
& K(\beta) = \nonumber \\
& \quad 4\sqrt{x}\frac{\beta^2 x^2+4\beta^2\left(\beta+1\right)x+4\beta^3+7\beta^2+4\beta+1}{\left(1+x\right)^2 \left[ \left(\beta+1\right)^2 + x \beta^2\right]^2},
\label{eq:K}
\end{align}
and $x$ is the only real root of the cubic equation
\begin{align}
& \beta^3 x^3 + \beta \left( 7\beta^2 + \beta -1 \right)x^2 + \left( 3\beta^3 + 2\beta^2 + 2\beta + 1\right) x \nonumber \\
& \qquad - \left( 3\beta^3 + 7\beta^2 + 5\beta + 1\right) = 0. 
\label{eq:x}
\end{align}
As can be seen from Eqs.~\eqref{eq:K} and~\eqref{eq:x}, $K(0)=1$, and $K(\beta)\rightarrow 0$ when $\beta\rightarrow1$.

\begin{figure}[!th]
\includegraphics[trim = 5mm 0 15mm 10mm, clip = true, width=\columnwidth]{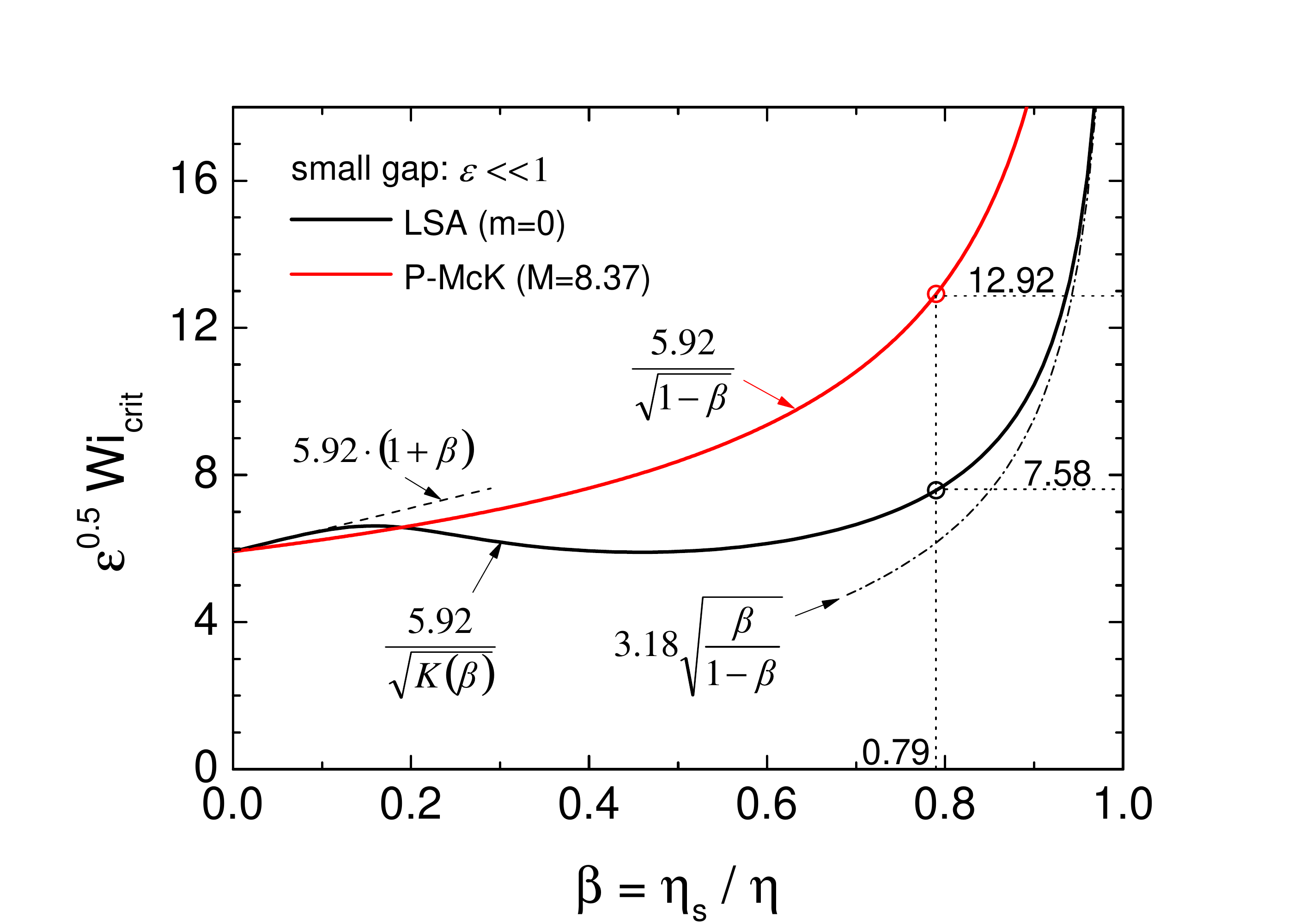}%
\caption{Onset of elastic instability according to the Pakdel-McKinley criterion \eqref{eq:PMcK_TC_OB_normalWi} (with $M=8.37$) and linear stability analysis \eqref{eq:PMcK_sg_LSA} (axisymmetric, $m=0$ \cite{Larson1990}) in the small-gap approximation, $\varepsilon \ll 1$.}%
\label{fig:sg_beta_to_1}%
\end{figure}

In the small-gap approximation, the result of the linear stability analysis, Eq.~\eqref{eq:PMcK_sg_LSA}, has a form similar to the Pakdel-McKinley condition, Eq.~\eqref{eq:PMcK_TC_OB_normalWi}: both equations predict a modified threshold $\sqrt{\varepsilon}Wi$ that is independent of $\varepsilon$. Eq.~\eqref{eq:PMcK_sg_LSA}, however, has a more complicated dependence on $\beta$ than the small-gap Pakdel-McKinley condition, Eq.~\eqref{eq:PMcK_TC_OB_normalWi}. In order to quantify this difference, we fix the constant $M$ in Eq.~\eqref{eq:PMcK_TC_OB_normalWi} to $M=5.92\sqrt{2}=8.37$ to match both equations for $\beta=0$, and plot their predictions as a function of $\beta$ in Figure \ref{fig:sg_beta_to_1}. While the overall shape of both curves is the same, the Pakdel-McKinley criterion predicts significantly higher values of the instability threshold. For example, for $\beta=0.79$, it gives $\sqrt{\varepsilon}Wi_\text{crit} \approx 12.92$ whereas the linear stability analysis predicts $\sqrt{\varepsilon}Wi_\text{crit} \approx 5.92/\sqrt{K(0.79)} = 7.58$.

\begin{figure}[!th]
\includegraphics[trim = 0 0 0 0, clip = true, width=\columnwidth]{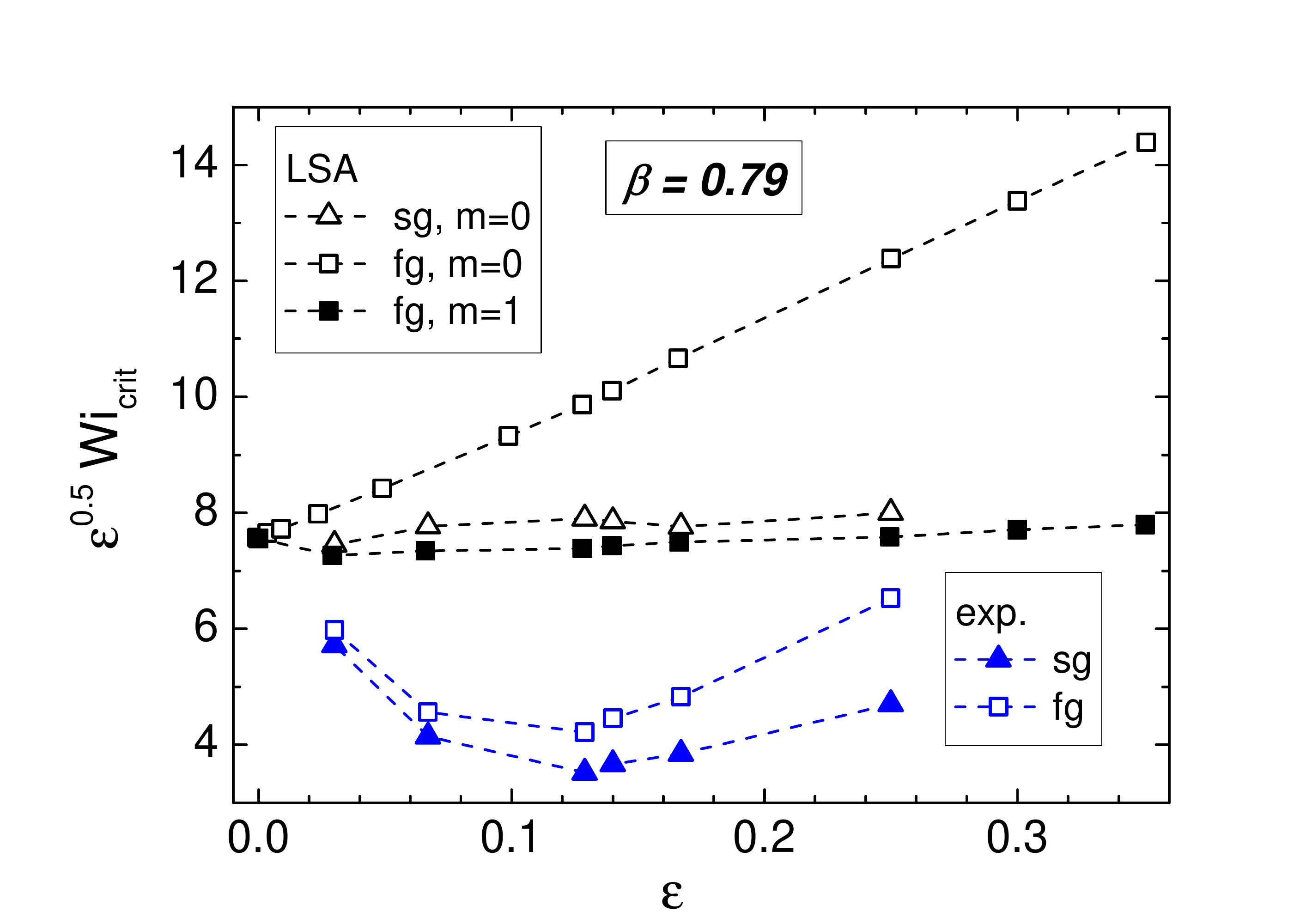}%
\caption{Results from various linear stability analyses (small gap (sg): $m=0$ \cite{Larson1990}; finite gap (fg): $m=0$ \cite{Shaqfeh1992}, $m=1$ \cite{Joo1994}) and experiments \cite{Muller1989} for the onset of linear visco-elastic instability of an Oldroyd-B fluid with $\beta = 0.79$. Data are extracted from the original publications.}%
\label{fig:motivation}%
\end{figure}

The small-gap approximation was relaxed by Joo and Shaqfeh \cite{Joo1994,Shaqfeh1992} who performed numerical linear stability analysis of Taylor-Couette flow explicitly taking into account the variation of the shear rate in the gap. 
Figure \ref{fig:motivation} shows their results for an Oldroyd-B fluid with $\beta=0.79$. For the axisymmetric mode, $m=0$, Joo and Shaqfeh compared numerical linear stability performed with the small- and finite-gap approximations, and observed that the finite-gap effects stabilise the flow. A similar observation was made earlier for the inertial instability of Newtonian Taylor-Couette flow \cite{Esser1996, Dutcher2007}. 

While until now we have only considered the axisymmetric mode of perturbation, non-axisymmetric modes, $m \geq 1$, are well known to be more unstable \cite{Joo1994, Avgousti1993}. In Fig.\ \ref{fig:motivation} we, therefore, also plot the linear stability threshold corresponding to the first non-axisymmetric mode ($m=1$) for $\beta=0.79$ within the finite-gap (fg) approximation. As can be seen from Fig.\ \ref{fig:motivation}, the $m=1$ neutral curve shows a very different geometric scaling on $\varepsilon$ compared to the axisymmetric ($m=0$) mode.

To our knowledge,  the only experimental study focusing on the explicit geometric scaling of the elastic instability threshold in Taylor-Couette can be found in the work of Larson, Shaqfeh and Muller \cite{Muller1989,Larson1990}. Their study used a Boger fluid with $\beta\approx0.79$ and, thus, can be directly related to the linear stability analysis presented above. 
The corresponding critical modified Weissenberg numbers are presented in Fig.\ \ref{fig:motivation}. It is evident that the experimental data exhibit significant discrepancies with the results of the linear stability analysis, both in absolute numbers and their dependence on the relative gap-size $\varepsilon$.\\

A resolution of this discrepancy was suggested by Sureshkumar and co-workers \cite{Al-Mubaiyedh1999,Al-Mubaiyedh2000}, who noted that the Boger fluid used by Larson, Shaqfeh and Muller \cite{Muller1989,Larson1990} should be sensitive to relatively small temperature variations in the sample. Since a part of the work spent on constant shearing of the fluid goes into viscous heating, Sureshkumar and co-workers \cite{Al-Mubaiyedh1999,Al-Mubaiyedh2000} argued that the material properties of the fluid should be shear-rate-dependent. They performed the corresponding linear stability analysis of the Oldroyd-B model coupled to the temperature equation, and found that for realistic thermal properties of the fluid the instability threshold is significantly lower than its constant temperature counterpart, and that the most unstable mode is axisymmetric ($m=0$).

\subsection{Results}
In this section we present the results of the linear stability analysis for our system. We linearise the equations of motion Eq.\eqref{eq:oldroyd-b} and Eqs.\eqref{eq:ns}-\eqref{eq:sPTT} around the base state, discretise them using the pseudospectral Chebyshev-tau method \cite{Canuto:book} and solve numerically the resulting generalised eigenvalue problem using Scientific Python \cite{scipy}.

As the first step, we compare predictions of our code against the existing results discussed above. We recalculate the critical Weissenberg number $Wi(\varepsilon)$ for the purely elastic ($Re=0$) Taylor-Couette flow of a UCM fluid ($\beta=0$) and an Oldroyd-B fluid with $\beta=0.79$; $\eta_p=\alpha = \lambda_p = 0$ in both cases. We consider axisymmetric ($m=0$) and the first non-axisymmetric ($m = 1$) modes.
\begin{figure}[!th]
\includegraphics[trim = 5mm 0mm 15mm 5mm, clip=true, width=\columnwidth]{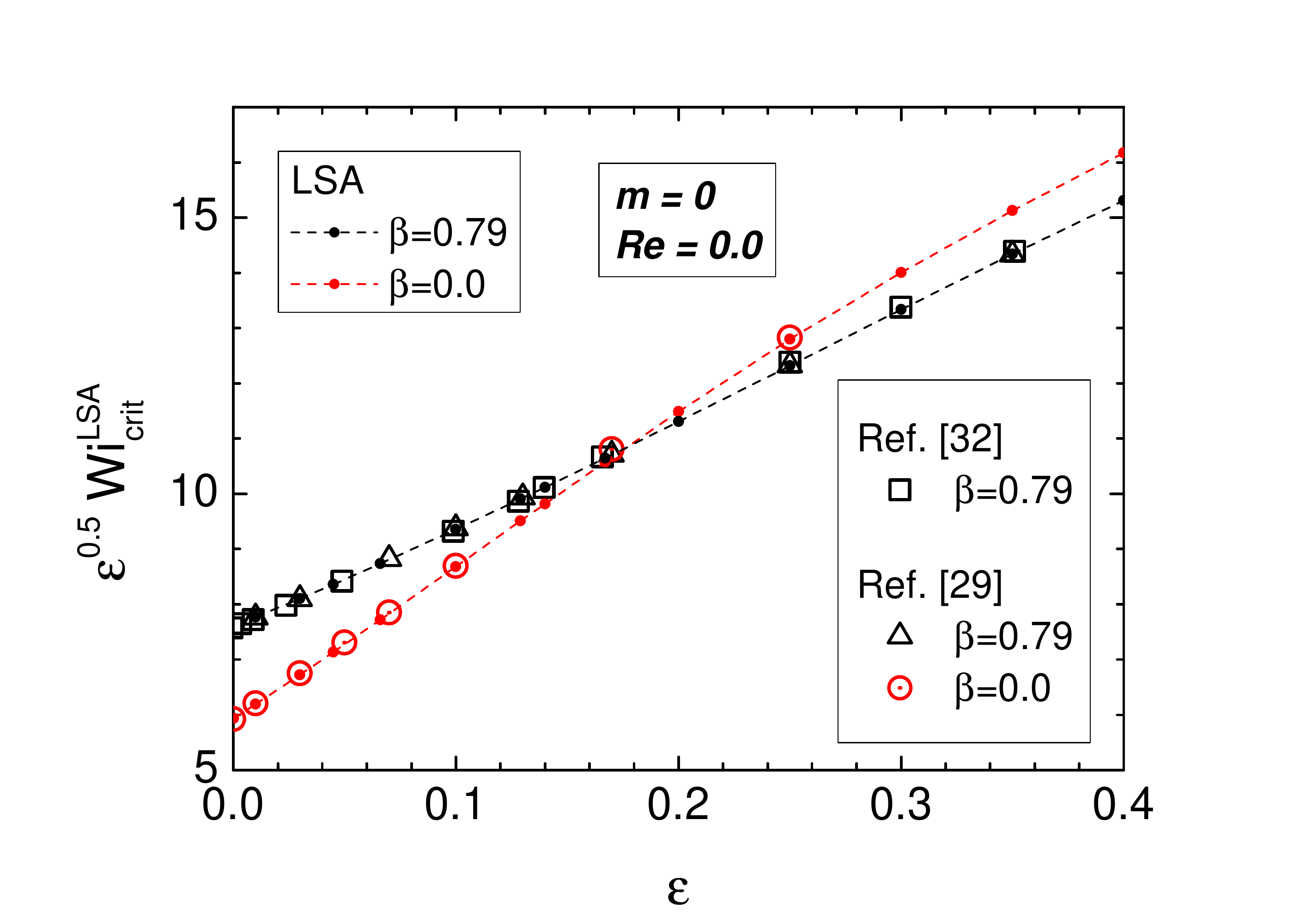}%
\caption{Results of our linear stability analysis of the purely elastic ($Re=0$) base Taylor-Couette flow of a Maxwell ($\beta=0$) and Oldroyd-B fluid ($\beta=0.79$) for small axisymmetric ($m=0$) disturbances. Our data reproduce the results published in Ref.\ \cite{Joo1994} and Ref.\ \cite{Shaqfeh1992}.} 
\label{fig:O-B_consistency}
\end{figure}

\begin{figure}
\includegraphics[trim = 5mm 0mm 15mm 5mm, clip=true, width=\columnwidth]{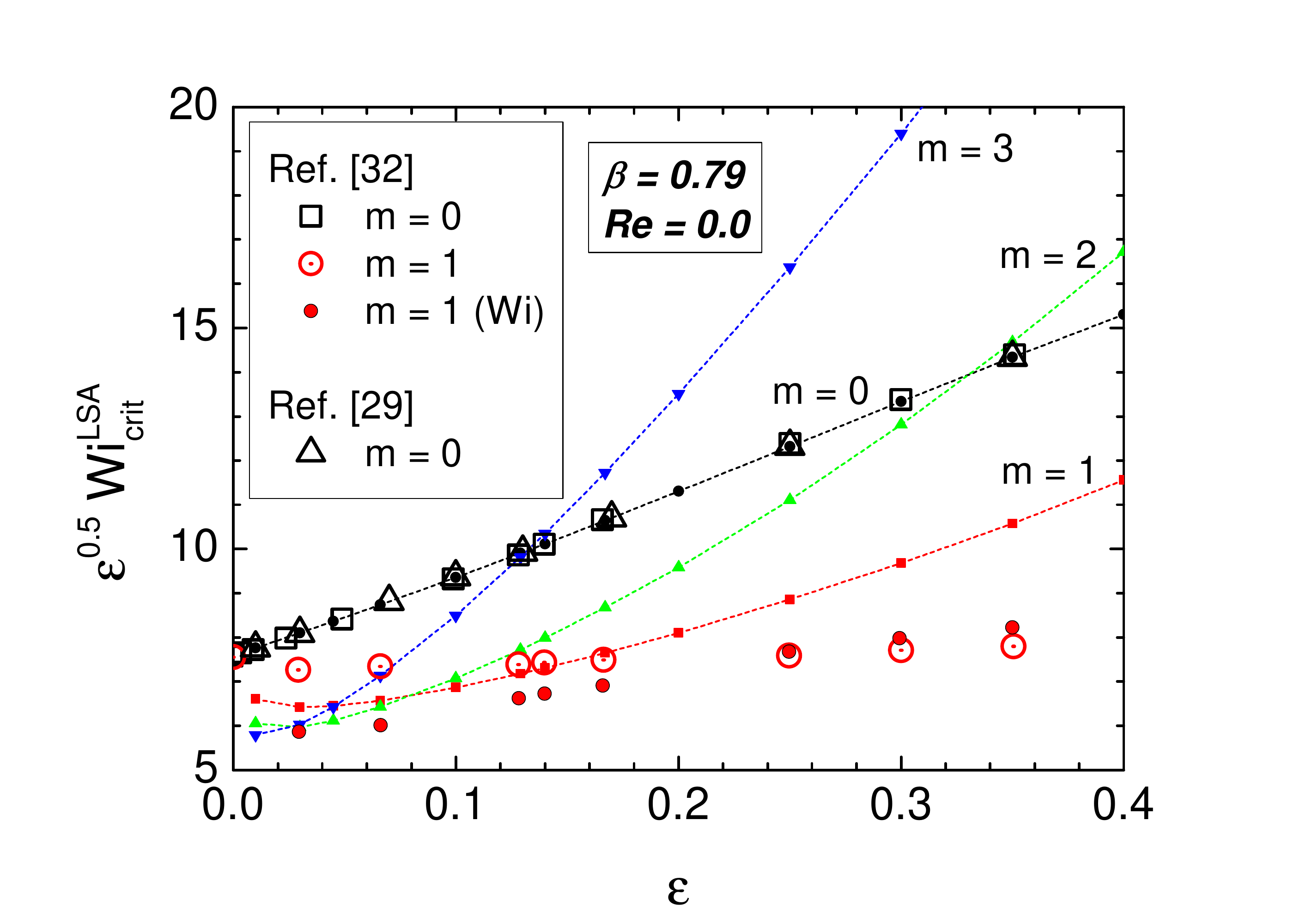}
\caption{Critical (modified) Weissenberg number as a function of the relative gap width. The referenced data are extracted from linear stability analysis of a Maxwell fluid ($\beta=0$) and an Oldroyd-B fluid ($\beta=0.79$), the latter for both, axisymmetric ($m=0$) as well as first non-axisymmetric mode ($m=1$) of disturbances \cite{Shaqfeh1992,Joo1994}. 
}
\label{fig:O-B_consistency_2}
\end{figure}

For $m=0$ in the small gap approximation \cite{Larson1990} our results are consistent with the data from the literature \cite{Shaqfeh1992,Joo1994}, see Fig.\ \ref{fig:O-B_consistency}. In contrast, our $m=1$ data significantly differs from the existing results (cf.\ Fig.\ \ref{fig:O-B_consistency_2}). We suspect a numerical error in the data of Ref.\ \cite{Joo1994} since we were also unable to convert these data for the critical Weissenberg number into the modified Weissenberg number given in the same publication (see Figs.12 and 13 in Ref.\ \cite{Joo1994}).

\begin{figure}[!th]
\includegraphics[trim = 0mm 2mm 0mm 5mm, clip=true, width=\columnwidth]{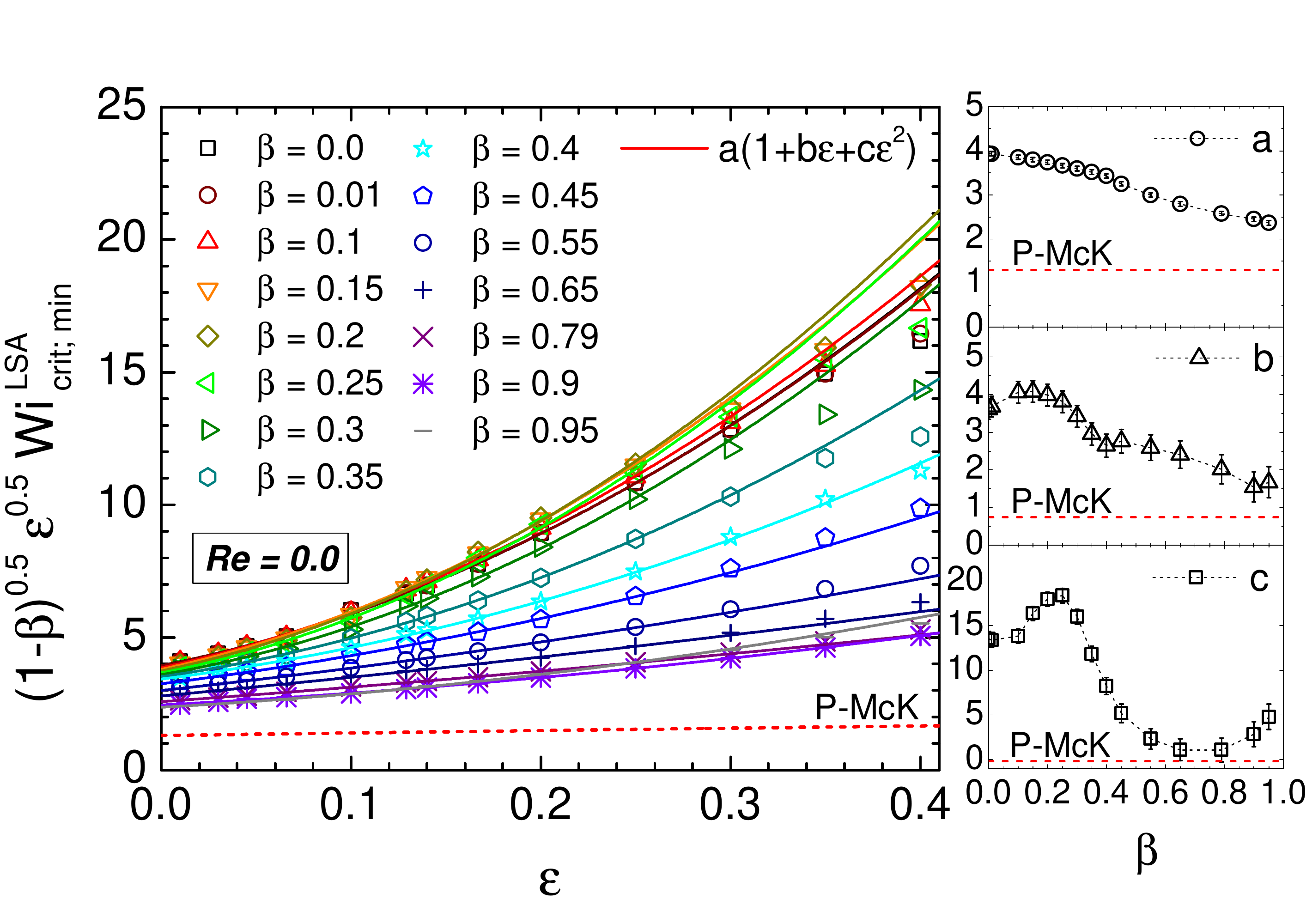}
\caption{Modified Weissenberg number for the most unstable mode as a function of relative gap width $\varepsilon$ for different values of relative viscosity $\beta=\eta_s/\eta_0$. The data are rescaled by a pre-factor $\sqrt{1-\beta}$ as suggested by the Pakdel-McKinley criterion \eqref{eq:PMcK_TC_OB_normalWi}. For values $\varepsilon \leq 0.35$, the data can be fairly fit by heuristic quadratic functions (lines, cf.\ eq.\ \eqref{eq:rescmodWi1}). The right panel shows the respective fit parameters $a,b$ and $c$, each compared with the prediction of the Pakdel-McKinley criterion \eqref{eq:rescmodWi2}.}
\label{fig:LSA_summary}
\end{figure}

\begin{figure}[!th]
\includegraphics[trim = -10mm 0mm 25mm 5mm, clip=true, width=0.95\columnwidth]{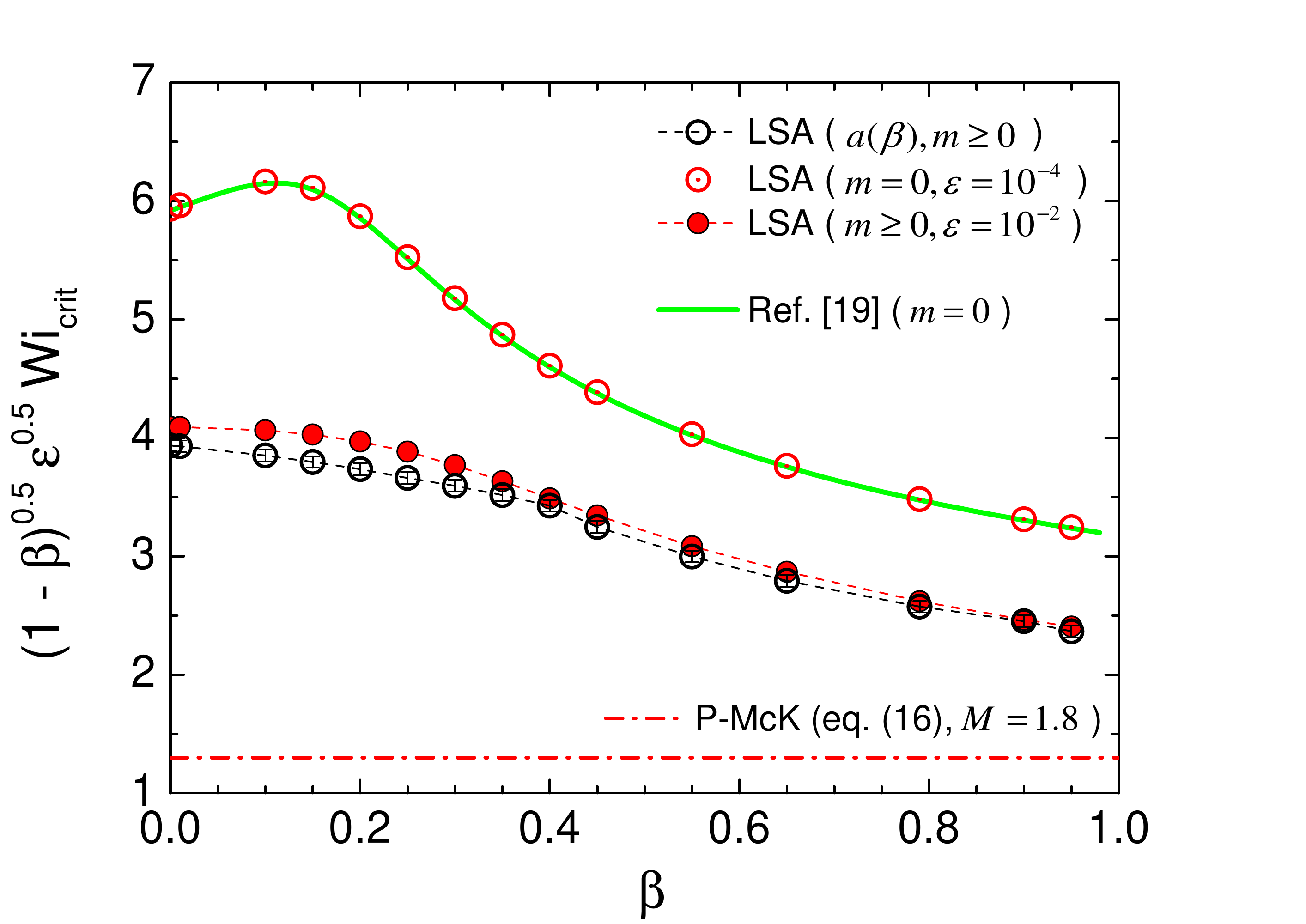}
\caption{Comparison of the different approaches for the rheological scaling of the onset of elastic instability in the small gap limit $\varepsilon \to 0$.}
\label{fig:a_vs_beta}
\end{figure}

As can be seen in Figure \ref{fig:O-B_consistency_2}, the $m=1$ mode is the most unstable one only for a limited range of gap widths, $\varepsilon \gtrsim 0.07$. For smaller $\varepsilon$, modes with higher azimuthal wavenumbers $m$ are more unstable, and the critical value of $m$ steadily increases upon decreasing the curvature. We interpret this as an indication that for vanishing curvature the system approaches the limit of plane Couette flow, a fact that is well known in Newtonian Taylor-Couette flow \cite{Faisst2000}.  Table \ref{tab:summary_Wi_crit} gives the critical Weissenberg number for various values of $\beta$, $\varepsilon$ and $m$. In the following, when talking about the critical onset for elastic instability, we will refer to the most unstable mode. The corresponding values are highlighted in boldface in Table \ref{tab:summary_Wi_crit} and presented in Fig.\ \ref{fig:LSA_summary}. For all values of $\beta$, the critical modified Weissenberg number obeys a heuristic quadratic dependency on the gap width $0 < \varepsilon \lesssim 0.3$:
\begin{equation}
\label{eq:rescmodWi1}
\sqrt{1-\beta}\sqrt{\varepsilon}Wi_\text{crit}^\text{LSA} = a(\beta) \Bigl(1 + b(\beta)\varepsilon + c(\beta)\varepsilon^2 \Bigr),
\end{equation}
where the functions $a(\beta),b(\beta),c(\beta)>0$ are presented in Fig.\ \ref{fig:LSA_summary}. In contrast, a Taylor expansion of the modified Weissenberg number according to the Pakdel-McKinley criterion, Eq.\eqref{eq:PMcK_fg_rheo}, predicts to second order
\begin{equation}
\label{eq:rescmodWi2}
\sqrt{1-\beta}\sqrt{\varepsilon}Wi_\text{crit} = \frac{M}{\sqrt{2}} \left(1 + \frac{3}{4}\varepsilon - \frac{5}{32}\varepsilon^2 \right) + \mathcal{O}(\varepsilon^3).
\end{equation}
It is obvious that the Pakdel-McKinley criterion and the linear stability analysis provide significantly different results. According to the former, the rescaled modified Weissenberg number $\sqrt{1-\beta} \sqrt{\varepsilon} Wi$ is independent of $\beta$ and in the small-gap limit it is equal to a constant, $M/\sqrt{2}$. Our experiments provide a value of $M = 1.8 \pm 0.3$. On the other hand, the linear stability analysis predicts that in the small-gap limit the rescaled modified Weissenberg number depends on $\beta$ since $a(\beta)$ is a decreasing function (see Figs.\ref{fig:LSA_summary} and \ref{fig:a_vs_beta}). We conclude that the Pakdel-McKinley criterion predicts a much weaker dependence on the gap width than the linear stability analysis since for any $\beta$ its coefficients of the correction terms $\mathcal{O}(\varepsilon)$ and $\mathcal{O}(\varepsilon^2)$ are much smaller than the functions $b(\beta)$ and $c(\beta)$ extracted from the linear stability analysis of the Oldroyd-B model.

Next we perform the linear stability analysis of our hybrid model in order to assess the influence of shear-thinning on the stability threshold and its scaling with the gap width. As already mentioned above, the proper way to compare the results of the linear stability analysis for both the Oldroyd-B and hybrid models against our experimental results, all three should be formulated in terms of the rheological Weissenberg number, defined in Section \ref{subject:rheology}. Since the base profile of our shear-thinning hybrid model differs from the Oldroyd-B one, we calculate it numerically, and use the maximum value of the velocity gradient, which is always at the inner cylinder, to define the rheological Weissenberg number for this model according to Eq.\eqref{eq:Wi_rheo_hybrid}.

\begin{figure}[!th]
\includegraphics[trim = -5mm 0 0 0, clip = true, width=\columnwidth]{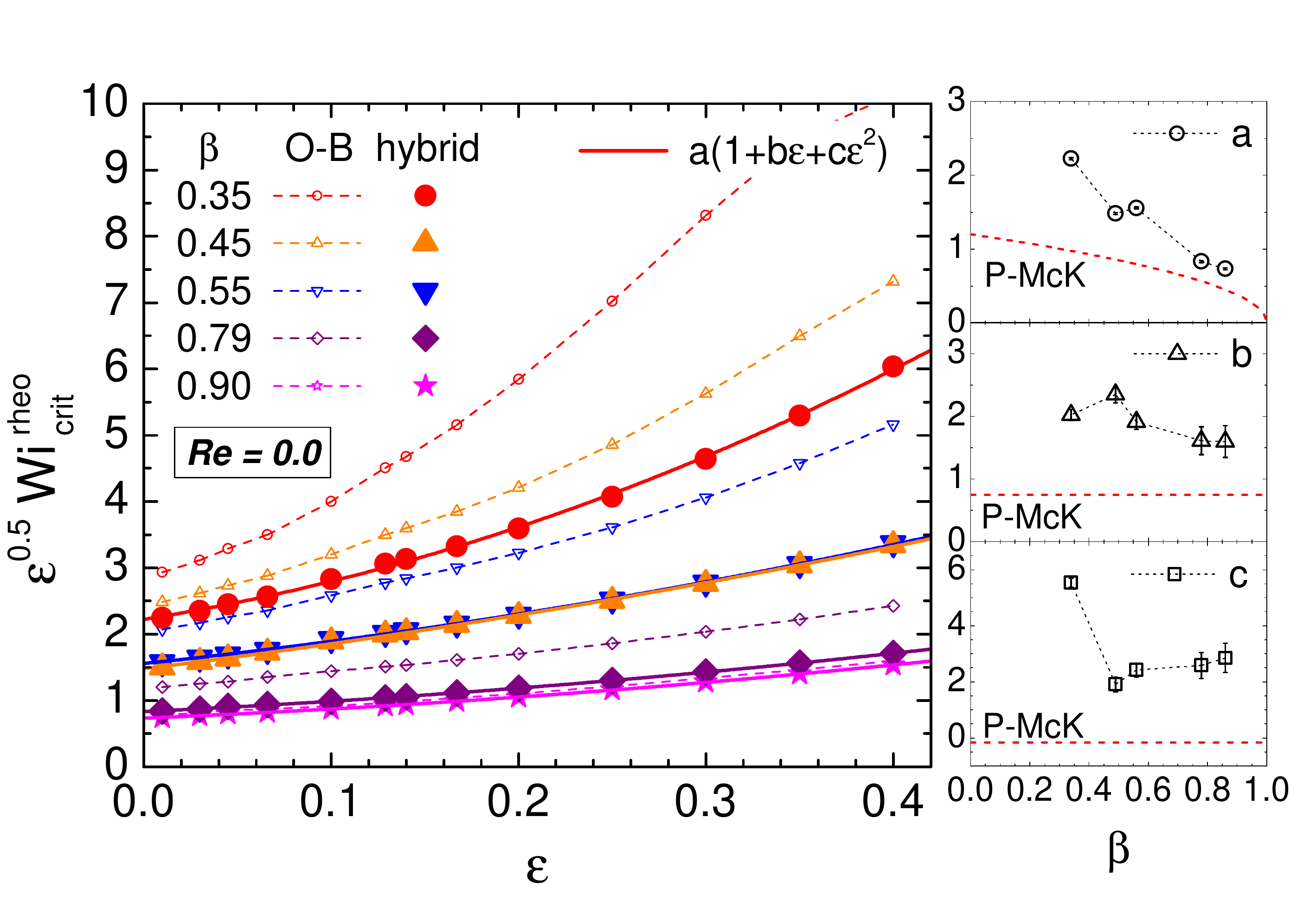}
\caption{Results of the linear stability analysis of two-mode hybrid model compared with results from the Oldroyd-B model only.}
\label{fig:LSA_summary_hybrid}
\end{figure}

\begin{figure}[!th]
\includegraphics[trim = 10mm 0 10mm 0, clip = true, width=\columnwidth]{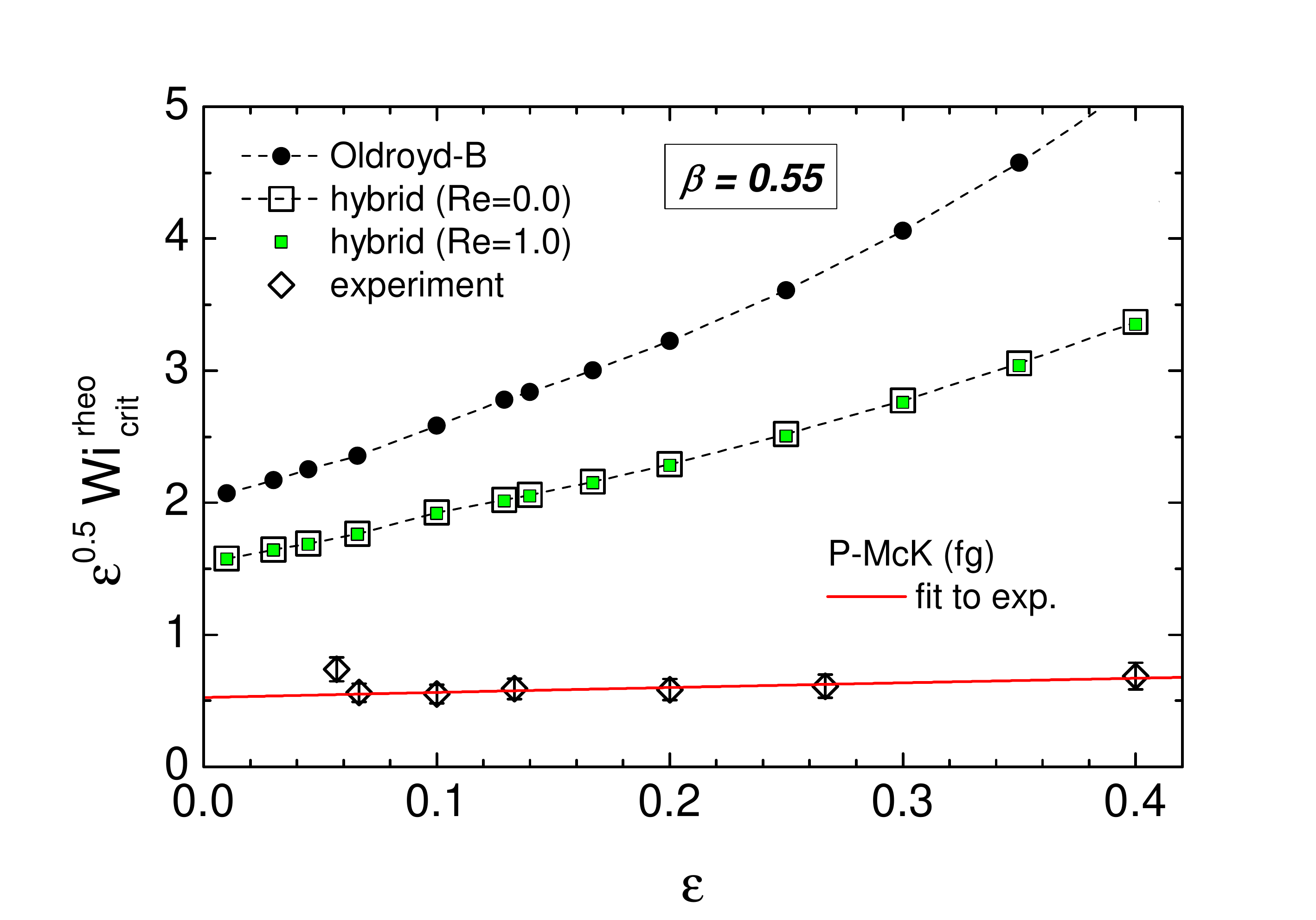}
\caption{Comparison of the different theoretical results on the onset of elastic instability compared with the experimental results for the P600$_\text{G80}$ ($\beta=0.55$) solution.}
\label{fig:LSA_summary_hybrid_2}
\end{figure}

In Fig.\ \ref{fig:LSA_summary_hybrid} we compare the linear stability analysis of the Oldroyd-B and the hybrid model for a range of $\beta = \eta_s/\eta_0$. The hybrid model clearly predicts a lower onset than the Oldroyd-B model.
Nevertheless, a significant discrepancy with the experimental data still remains as illustrated in Fig.\ \ref{fig:LSA_summary_hybrid_2} for $\beta = 0.55$. We have checked that this problem is not cured by the addition of a small amount of inertia. In Fig.\ \ref{fig:LSA_summary_hybrid_2} we also present the results of the linear stability analysis of the hybrid model for the Reynolds number $Re = 1.0$. This value is an estimate from above of the maximal Reynolds number reached in our experiments. The instability threshold in the presence of a small amount of inertia is almost identical to the purely elastic case and the discrepancy with the experimental observations remains. 

\section{LSA of the non-isothermal hybrid model}
\label{subsec:LSA_noniso}
In the previous section we have demonstrated that neither the Oldroyd-B nor the shear-thinning hybrid model can predict the experimentally observed instability threshold. As we already mentioned above, this problem is reminiscent of the discrepancy between the instability threshold for a Boger fluid in a small-gap Taylor-Couette setup observed by Larson, Shaqfeh and Muller \cite{Larson1990}, and the predictions of the linear stability analysis for the Oldroyd-B model. This discrepancy was resolved in \cite{Al-Mubaiyedh1999,Al-Mubaiyedh2000,White2000,White2003}, where it was argued that constant shearing by the base flow and small perturbations result in local viscous heating of the fluid rendering its mechanical properties spatially inhomogeneous. Since Boger fluids are typically based on rather viscous Newtonian solvents \cite{James2009}, they are prone to significant viscous heating. For realistic thermal properties of Boger fluids, Sureshkumar \emph{et al.} \cite{Al-Mubaiyedh1999,Al-Mubaiyedh2000} found that the instability threshold is significantly lower than its constant-temperature counterpart. In this section we examine whether the same argument resolves a similar discrepancy for our dilute polymer solutions that are significantly less viscous than the Boger fluid employed by Larson, Shaqfeh and Muller \cite{Larson1990}.

First we note that there are reasons to expect temperature gradients in our setup. The inner cylinder of our Taylor-Couette cell is mounted on the driving motor which is at ambient temperature ($\approx 23^{\circ}$C), and the upper part of the cylinder is not immersed into the fluid (see Fig.\ref{fig:geometry}). It is, therefore, reasonable to assume that there is a temperature difference between the inner cylinder and the outer beaker which is in contact with the surrounding thermal bath fixed at $T=T_0$. This difference should be especially significant for the measurements at $T_0= 10\,^{\circ}$C, and we expect $T(R_1) > T(R_2)=T_0$. We have verified this assumption by measuring the steady state temperature difference between the thermal bath and the temperature of the inner cylinder in a particular setup, measured by a temperature sensor mounted in an bore hole close to the surface of the cylinder. The observed equilibrium temperature difference is approximately $\Delta T = T(R_1) - T_0 \lesssim 0.2\,^{\circ}$C.

To include the effect of viscous heating in our hybrid model, we turn to one of the simplest forms of the kinetic theory for dilute polymer solutions -- a suspension of non-interacting Hookean dumbbells \cite{bird2}. Within this theory, the polymeric contribution to the stress tensor $\bm{\tau}$ is given by
\begin{gather}
\bm{\tau} = n H \langle \bm{Q}\bm{Q}\rangle - n k_B T \bm{\delta},
\end{gather}
and, simultaneously, by
\begin{gather}
\bm{\tau} = - \frac{n \zeta}{4} \ucd{\langle \bm{Q}\bm{Q}\rangle}.
\end{gather}
Here, $\bm{Q}$ is the end-to-end vector of a polymer chain, $n$ is the number density of polymer chains, $H$ is the Hookean spring constant, $\zeta$ is the coefficient of friction between the polymer chain and the solvent, $k_B$ is the Boltzmann constant, and the ensemble average is performed over the equilibrium distribution function $\Psi(\bm{Q})$ \cite{bird2,Morozov2015}. Combining these two equations yields the Oldroyd-B model. To account for spatial temperature variations, we observe that the Hookean elastic constant $H$ has entropic origins \cite{DoiEdwards}, and, therefore, $H\propto T$. The friction coefficient $\zeta$, on the other hand, is proportional to the viscosity of the Newtonian solvent suspending polymer chains, that typically obeys the Arrhenius-type law
\begin{gather}
\zeta \sim \eta_s (T) \sim e^{\frac{E_a}{R T}},
\label{eq:Arrhenius}
\end{gather}
where $E_a$ is the activation energy, and $R$ is the gas constant \cite{Tanner1985}. Taking into account this explicit temperature dependence, and assuming that the temperature can vary in space and time, we combine the above expressions for the stress tensor to obtain a non-isothermal version of the Oldroyd-B model, Eq.\eqref{eq:oldroyd-b},
\begin{gather}
\bm{\tau}_u + \lambda_u^{0} e^{\nu \left( \frac{T_0}{T} - 1\right)} \frac{T_0}{T} 
\left[ \ucd{\bm{\tau}_u}  - \bm{\tau}_u \frac{1}{T} \left( \frac{\partial}{\partial t} + \vec{v} \cdot \nabla \right) T\right] \nonumber \\
= \eta_u^{0} e^{\nu \left( \frac{T_0}{T} - 1\right)}\left(\nabla\vec{v} + \nabla\vec v^\dagger\right).
\label{eq:non-iso-oldroyd-b}
\end{gather}
Here, $T=T(\bm{r},t)$ is a local value of the temperature, $T_0$ is the reference temperature of the fluid at rest, $\nu = \frac{E_a}{R T_0}$ is the dimensionless activation energy, and we have identified the usual kinetic-theory expressions for the polymer viscosity and relaxation time at the reference temperature
\begin{gather}
\lambda_u^{0} = \frac{\zeta(T_0)}{4 H(T_0)}, \\
\eta_u^{0} = \lambda_u^{0} n k_B T_0.
\end{gather}

In a similar fashion, the non-isothermal version of the sPTT model, Eq.\eqref{eq:sPTT}, is given by
\begin{gather}
\bm{\tau}_p \left(1 + \alpha\frac{\lambda_p^{0}}{\eta_p^{0}}\frac{T_0}{T} \text{tr}\bigl(\bm{\tau}_p\bigr)\right) \nonumber \\
+ \lambda_p^{0} e^{\nu \left( \frac{T_0}{T} - 1\right)} \frac{T_0}{T} 
\left[ \ucd{\bm{\tau}_p}  - \bm{\tau}_p \frac{1}{T} \left( \frac{\partial}{\partial t} + \vec{v} \cdot \nabla \right) T\right] \nonumber \\
= \eta_p^{0} e^{\nu \left( \frac{T_0}{T} - 1\right)}\left(\nabla\vec{v} + \nabla\vec v^\dagger\right).
\end{gather}
The temperature field is assumed to obey the advection-diffusion equation \cite{ll87}
\begin{gather}
\rho\,c_p \left( \frac{\partial T}{\partial t} + \vec{v}\cdot\nabla T\right) = \kappa \nabla^2 T + \Sigma'_{ij}\frac{\partial v_i}{\partial x_j},
\label{tempeq}
\end{gather}
where $c_p$ and $\kappa$ are the heat capacity and thermal conductivity of the fluid, respectively, and the non-isothermal deviatoric stress $\bm{\Sigma}'$ is given by
\begin{gather}
\label{eq:sigmaprime}
\bm{\Sigma}' = \eta_s(T_0) e^{\nu \left( \frac{T_0}{T} - 1\right)}  \left(\nabla\vec{v} + \nabla\vec v^\dagger\right) + \bm{\tau}_u + \bm{\tau}_p.
\end{gather}
The velocity field satisfies Eqs.\eqref{eq:ns} and \eqref{eq:incomp} with $\bm{\Sigma} = \bm{\Sigma}' - p\,\bm{\delta}$. Finally, the boundary conditions for the temperature field are set by our experimental setup: the outer cylinder is in contact with the heat bath at fixed temperature $T_0$, while the inner cylinder can have a different temperature, which we do not control. Therefore, we set
$T(R_2) = T_0$ and $T(R_1) = T_0+\Delta T$. As mentioned above, we measured $\Delta T$ to be around $0.2K$.

The non-isothermal hybrid model presented above contains three parameters that need to be determined before it can be used in a linear stability analysis: the thermal conductivity, the heat capacity and the activation energy. In what follows, we restrict our discussion to the P600$_\text{G80}$ solution (see Table \ref{tab:polymersolutions}); other solutions show qualitatively similar behaviour (see \cite{Schafer2013} for more detail). To estimate the value of $\kappa$, we used the study by Broniarz-Press and Pralat \cite{Broniarz-Press2009} who systematically investigated the thermal conductivity of various Newtonian and non-Newtonian liquids, including high-molecular-weight polyacrylamide solutions (Separan) relevant for the present work. Broniarz-Press and Pralat \cite{Broniarz-Press2009} observed that the thermal conductivity is practically independent of the shear rate, and using their empirical equation we estimate it to be $\kappa \approx 0.7\, W\, m^{-1}\, K^{-1}$. The specific heat capacity of the P600$_\text{G80}$ solution was measured using a differential scanning calorimeter DSC Q2000, and found to be $c_p\approx 2.7\, kJ\, kg^{-1}\, K^{-1}$. The activation energy $\nu$ of the P600$_\text{G80}$ solution was estimated by fitting the temperature dependence of its solvent viscosity to an Arrhenius-type law, Eq.\eqref{eq:Arrhenius},
\begin{equation}
\frac{\eta_s(T)}{\eta_s(T_0)} = e^{\nu\left(\frac{T_0}{T}-1\right)}.
\label{eq:Arrheniusfit}
\end{equation}
Using a pseudo-empirical formula by Cheng \cite{Cheng2008} for the viscosities of the glycerol solvents, we obtain $\nu\approx 20.27$. A similar value was recently reported by Traore \emph{et al.} \cite{Traore2015} for a sucrose-water PAAm solution, similar to our P80$_\text{S64}$ and P150$_\text{S65.6}$ solutions, see Table \ref{tab:polymersolutions}; see also Abed \emph{et al.} \cite{Abed2016}. Note, however, that Traore \emph{et al.} \cite{Traore2015} observed that the activation energy for the relaxation time was five times larger than the viscosity activation time, which is not captured within our theory based on the kinetic theory of dilute polymer solutions.

The results of the linear stability analysis of our non-isothermal hybrid model are reported in terms of two dimensionless numbers: the Peclet number,
\begin{align}
Pe = \frac{\rho\,c_p\dot{\gamma}_{max}d^2}{\kappa},
\end{align}
which is the ratio of the thermal diffusion timescale and the convective timescale, and the Nahme number,
\begin{align}
Na = \frac{\eta(T_0) (\dot{\gamma}_{max} d)^2}{\kappa T_0},
\end{align}
which compares the viscous heating and the convective timescales. Here, $\dot{\gamma}_{max}$ is the maximum shear rate of the base velocity profile in our non-isothermal hybrid model. Using the values of the viscous heating parameters estimated above, the Peclet and Nahme numbers achieved in our experiments are $Pe \le 400$ and $Na \le 1\cdot 10^{-5}$.

\begin{figure}[!th]
\centering
\includegraphics[trim = 10mm 0mm 25mm 15mm, clip=true, width=\columnwidth]{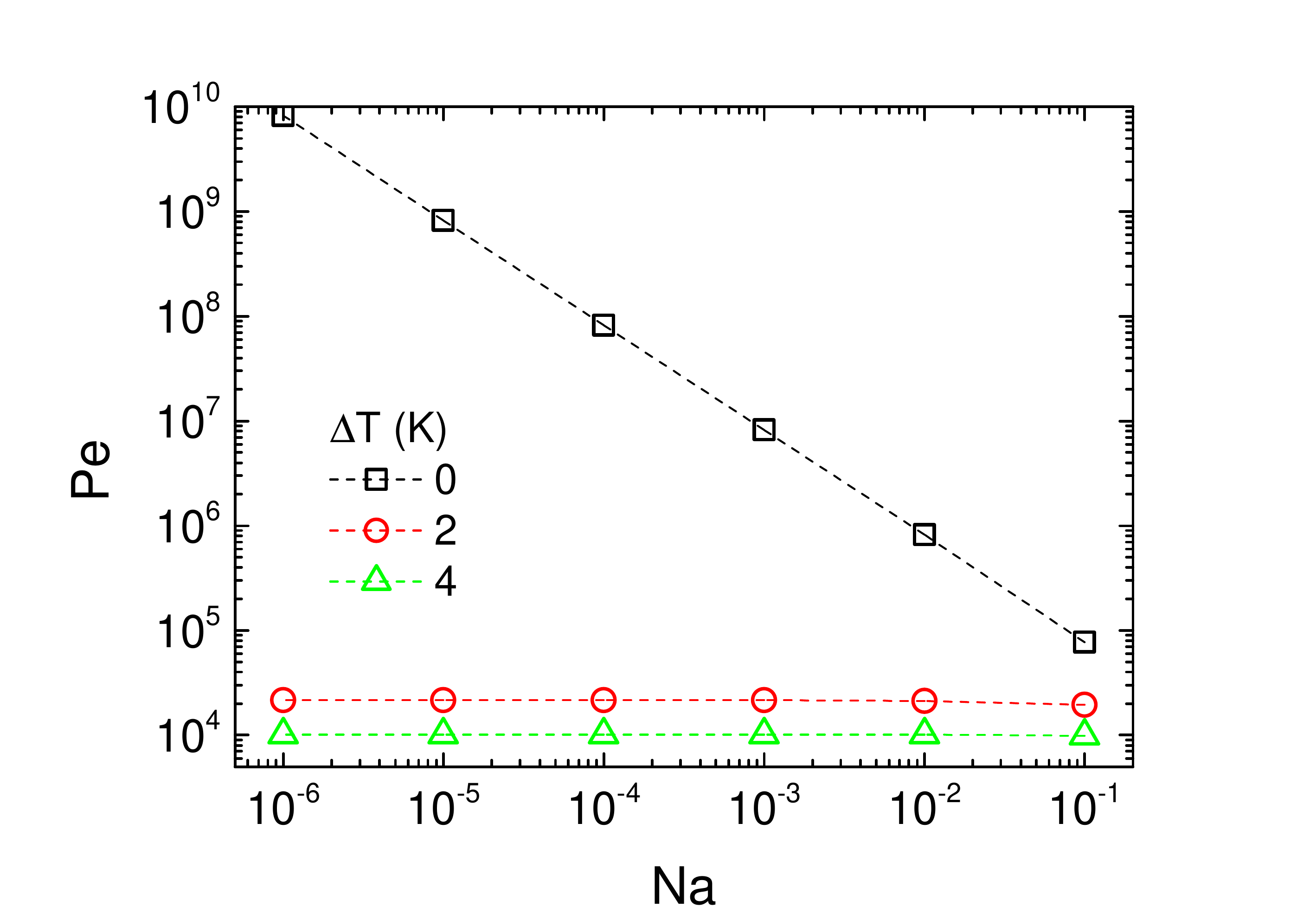}%
\caption{Nahme and Peclet numbers required for the critical modified Weissenberg number to be $\sqrt{\varepsilon} Wi = 0.65$ for the P600$_\text{G80}$ solution at $\varepsilon = 0.4$.}
\label{fig:Pe_vs_Na}
\end{figure}

Performing a linear stability analysis of the non-isothermal hybrid model with these values of the Peclet and Nahme numbers and setting $\Delta T=0.2K$, we observe only negligible difference in the values of the critical Weissenberg number as compared to the isothermal model discussed in the previous Section. This situation persists until $\Delta T>2K$, which is about ten times larger than the temperature difference between the inner and outer cylinders that we measured in our experiments. A more detailed analysis of the interplay between the thermodynamic parameters of the fluid and the onset of elastic instabilities reveals a much stronger impact of the Peclet number on the critical value of $Wi$ compared to the Nahme number. In Fig.\ref{fig:Pe_vs_Na} we force the critical rheological Weissenberg number from the linear stability analysis to agree with the experimental value $\sqrt{\varepsilon} Wi_c^\text{rheo}=0.65$ at $\varepsilon=0.4$, and record the Peclet and Nahme numbers that are necessary to achieve that. When there is no temperature difference between the inner and outer cylinders, $\Delta T = 0K$, $Pe$ has to be about $10^9$ for $Na=10^{-5}$ to observe an instability at the experimental value of $\sqrt{\varepsilon} Wi_c^\text{rheo}$. For higher temperature differences $\Delta T = 2K$ and $\Delta T = 4K$, the required Peclet number is approximately independent of the Nahme number, and is equal to $Pe = 21550$ and $Pe = 10150$, correspondingly.

\begin{figure}[!t]
\centering
\includegraphics[trim = 10mm 0mm 25mm 15mm, clip=true, width=\columnwidth]{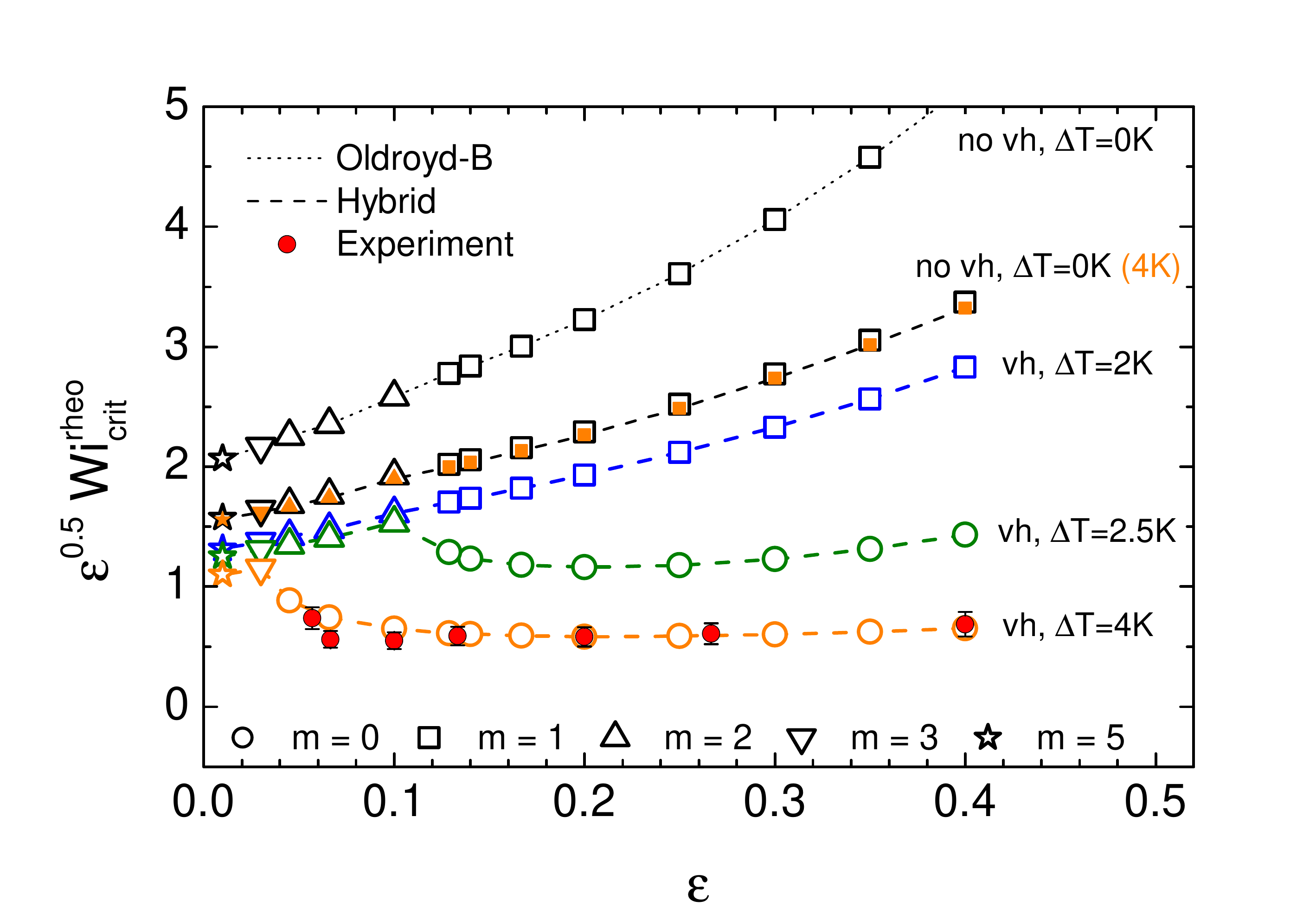}%
\caption[Effect of viscous heating in combination with a fixed temperature difference between the inner and outer wall based on the hybrid model]{Effect of viscous heating in combination with a fixed temperature difference between the inner and outer wall $\Delta T$ based on the hybrid model. While the activation energy $\nu$ as well as the Nahme number $Na$ are assumed to be in reasonable agreement with the estimates, the Peclet number $Pe$ is more than one magnitude higher than estimated. Representatively, the value $\varepsilon=0.3$ is analysed in more detail in Fig. \ref{fig:energetics}.}%
\label{fig:temperature}
\end{figure}

Although $Pe=10150$ and $\Delta T = 4K$ are very different from the estimates for our polymer solutions, we use these values in the further analysis to demonstrate that viscous heating has a potential to significantly alter the scaling of the critical Weissenberg number with the gap width. Figure \ref{fig:temperature} illustrates the effect of the interplay between the temperature difference $\Delta T$ and viscous heating for $Na=10^{-5}$ and $Pe = 10150$; other parameters correspond to the P600$_\text{G80}$ solution. In the absence of viscous heating, i.e. the last term in Eq.\eqref{tempeq}, the prediction of the linear stability analysis strongly disagrees with the experimental data for both the Oldroyd-B and the hybrid model, regardless of $\Delta T$. In the presence of viscous heating, the geometric scaling depends strongly on the value of the temperature difference, and changes significantly for $\Delta T > 2K$: the experimental data, together with the scaling predicted by the Pakdel-McKinley criterion, can now be reproduced by the non-isothermal linear stability analysis. 

\begin{figure}[h]
\centering
\includegraphics[trim = 0mm 0mm 10mm 15mm, clip=true, width=\columnwidth]{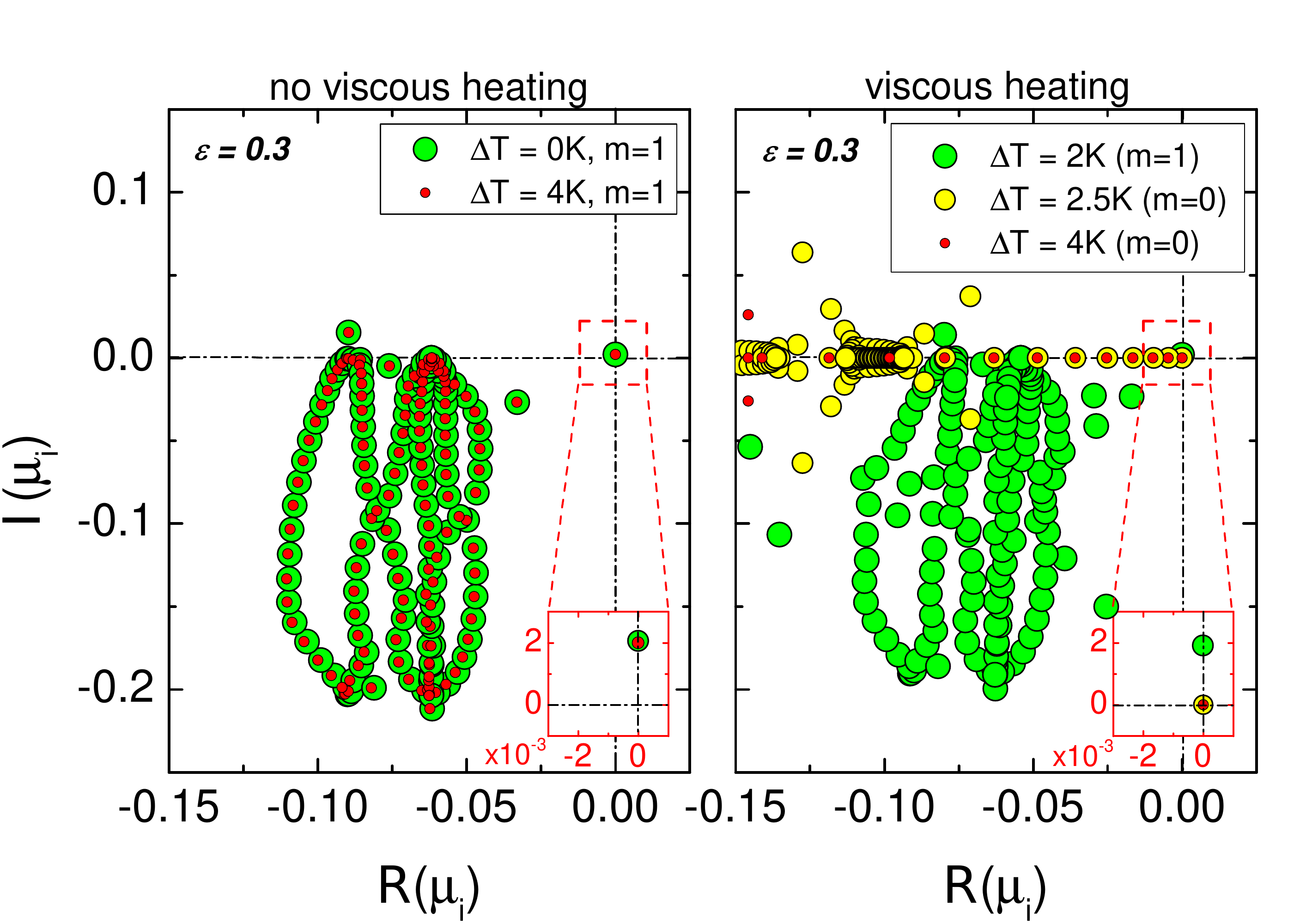}%
\caption{Eigenvalue spectrum for $\varepsilon = 0.3$ with and without viscous heating for various values of the temperature difference $\Delta T$. Thermodynamic parameters used in the viscous heating calculations are: $\nu = 20.27, Na = 1.4\cdot 10^{-5}, Pe = 10150$, and the hybrid model parameters correspond to the rheology of the P600$_\text{G80}$ solution ($\beta=0.55,\beta_1=0.18,\alpha=17.7,\lambda_r=1.53$).}
\label{fig:energetics}
\end{figure}

Furthermore, for $\Delta T > 2K$ and significantly large gap widths, the nature of the most unstable mode changes, in agreement with Al-Mubaiyedh \emph{et al.} \cite{Al-Mubaiyedh1999,Al-Mubaiyedh2000}. To illustrate this, in Figure \ref{fig:energetics} we plot the eigenvalue spectra for a given geometry $\varepsilon = 0.3$ and the rheological parameters corresponding to the P600$_\text{G80}$ solution; as above, $Na=10^{-5}$ and $Pe = 10150$. Without viscous heating (\textit{left panel}), the most unstable mode is non-axisymmetric ($m=1$) and oscillatory for both $\Delta T = 0K$ (cf.\ figure \ref{fig:temperature}) and $\Delta T=4K$. 

In contrast, in the presence of  viscous heating (\textit{right panel}), the behaviour changes upon an  increase of the temperature difference: at $\Delta T=2K$, the most unstable mode is still non-axisymmetric and oscillatory (green circles), while at $\Delta T=2.5K$ (yellow circles), the most unstable mode turns to be axisymmetric and stationary in time. At $\Delta T=4K$ (red circles), the spectrum is further changed but the most unstable mode is not affected.

\section{Conclusion}
The goal of this work was to study the influence of curvature on the purely elastic flow instability in the Taylor-Couette geometry. We performed systematic studies of the onset of the first instability in Taylor-Couette cells of various radii and observed that the onset shifts to larger critical Weissenberg numbers for increasing curvature. 
Simultaneously, we found that the intensity of the unstable flow decreases.
These observations are consistent with the fact that in this limit the geometry approaches plane Couette flow, which is known to be linearly stable for Oldroyd-B \cite{Gorodtsov1967} and FENE-P \cite{Arora2005} models.

We used our experimental results to verify the validity of the Pakdel-McKinley criterion across a wide range of curvatures. We found that when modified to take into account the finite gap width and the shear-thinning nature of our solutions, the Pakdel-McKinley criterion reproduces the experimentally observed critical Weissenberg numbers fairly well. Summarising the analysis of Section \ref{seq:PM}, the Pakdel-McKinley condition consistent with our data is given by
\begin{equation}
\nonumber
Wi^\text{rheo}_\text{crit} = (1.8 \pm 0.3)\sqrt{(0.96 \pm 0.05) - \beta^+} \sqrt{\frac{(\varepsilon + 1)^2}{\varepsilon+2}},
\end{equation}
where $ \beta^+$ is defined before Eq.\eqref{Meqtmp}. This scaling is consistent with the results of Groisman and Steinberg \cite{Groisman1998}, who investigated the onset of disordered oscillations for different polymer solutions in a single Taylor-Couette geometry, although the value of the constant $M$ found in that work differs from ours.

Surprisingly, we find significant discrepancies between the experimental data and the results of the linear stability analysis of both the Oldroyd-B model and a hybrid, Oldroyd-B-sPTT two-mode fluid designed to accurately reproduce rheological properties of our polymer solutions. These discrepancies comprise the numerical values of the critical Weissenberg number and its scaling with the dimensionless gap size $\varepsilon$. We checked that inclusion of small amounts of inertia, consistent with our experimental values, did not solve this problem. We also incorporated the effects of viscous heating into the hybrid model and demonstrated that it did not resolve the disagreement with the experimental data for realistic values of the thermodynamic parameters of the solutions and external temperature gradients in our setup. Intriguingly, the linear stability thresholds can be made to agree with the experimental data albeit for rather unrealistic values of the Peclet number and the temperature difference between the inner and outer cylinders.

There are several potential sources of the disagreement between our experiments and the linear stability analysis. While the hybrid model accurately reproduces the steady-shear rheology, it is possible that the linear instability threshold is sensitive to its unsteady behaviour and extensional properties, which we do not match. Another possibility is that the Ekman vortices, generated next to the upper surface and the bottom of the beaker affect the instability onset. Thus, it is even more remarkable that the simple Pakdel-McKinley condition agrees very well with the experimental data despite these potential problems.

\section{Acknowledgement}
CW an CS acknowledge support from the DFG (WA 1336/5-2). AM acknowledges support from EPSRC (grant number EP/I004262/1).

\begin{landscape}
\begin{table}[htbp]
	\renewcommand{\arraystretch}{0.5}
	\footnotesize
  \centering
    \begin{tabular}{cccccccccccccccccc}
		\toprule
    \multirow{2}[0]{*}{$\epsilon$} & \multirow{2}[2]{*}{$m$} & \multirow{2}[2]{*}{$\beta = $}& \multirow{2}[2]{*}{\textbf{0.00}} & \multirow{2}[2]{*}{\textbf{0.01}} & \multirow{2}[2]{*}{\textbf{0.10}} & \multirow{2}[2]{*}{\textbf{0.15}} & \multirow{2}[2]{*}{\textbf{0.20}} & \multirow{2}[2]{*}{\textbf{0.25}} & \multirow{2}[2]{*}{\textbf{0.30}} & \multirow{2}[2]{*}{\textbf{0.35}} & \multirow{2}[2]{*}{\textbf{0.40}} & \multirow{2}[2]{*}{\textbf{0.45}} & \multirow{2}[2]{*}{\textbf{0.55}} & \multirow{2}[2]{*}{\textbf{0.65}} & \multirow{2}[2]{*}{\textbf{0.79}} & \multirow{2}[2]{*}{\textbf{0.90}} & \multirow{2}[2]{*}{\textbf{0.95}} \\
&&&&&&&&&&&&&&&&\\
      \midrule
     & \multicolumn{1}{c}{0} &  & \textbf{25.57} & \textbf{26.15} & 33.09 & 36.09 & 34.39 & \textbf{30.44} & \textbf{27.07} & \textbf{24.67} & \textbf{23.02} & 21.93 & 20.91 & 21.17 & 24.20 & 32.53 & 44.65 \\
    \textbf{0.4} & \multicolumn{1}{c}{1} & $Wi_c=$ & 27.17 & 27.30 & \textbf{29.25} & \textbf{31.24} & 33.80 & 33.59 & 30.64 & 27.08 & 23.92 & \textbf{21.04} & \textbf{18.14} & \textbf{16.91} & \textbf{18.27} & \textbf{25.29} & \textbf{39.13} \\
          & \multicolumn{1}{c}{2} & & 30.85 & 30.87 & 31.40 & 31.88 & \textbf{32.37} & 32.56 & 31.73 & 29.89 & 27.81 & 25.96 & 23.47 & 22.77 & 26.43 & 42.92 & 77.69 \\
          \midrule
     & \multicolumn{1}{c}{0} & & 25.57 & 26.15 & 33.09 & 36.09 & 34.39 & 30.44 & \textbf{27.07} & \textbf{24.67} & 23.02 & 21.93 & 20.91 & 21.18 & 24.23 & 32.60 & 44.78 \\
     \textbf{0.35} & \multicolumn{1}{c}{1} & $Wi_c=$ & \textbf{25.26} & \textbf{25.39} & \textbf{27.20} & \textbf{29.05} & 31.42 & 31.24 & 28.49 & 25.18 & \textbf{22.26} & \textbf{19.96} & \textbf{17.19} & \textbf{16.32} & \textbf{17.88} & \textbf{24.71} & \textbf{37.87} \\
          & \multicolumn{1}{c}{2} & & 28.67 & 28.70 & 29.19 & 29.64 & \textbf{30.10} & \textbf{30.28} & 29.50 & 27.79 & 25.86 & 24.14 & 21.83 & 21.24 & 24.81 & 40.18 & 72.35 \\
           \midrule
      & \multicolumn{1}{c}{0} & & 25.57 & 26.15 & 33.09 & 36.09 & 34.39 & 30.44 & 27.07 & 24.67 & 23.03 & 21.93 & 20.94 & 21.23 & 24.35 & 32.81 & 45.09 \\
     \textbf{0.3} & \multicolumn{1}{c}{1} & $Wi_c=$ & \textbf{23.43} & \textbf{23.54} & \textbf{25.22} & \textbf{26.94} & 29.14 & 28.96 & \textbf{26.42} & \textbf{23.35} & \textbf{20.69} & \textbf{18.68} & \textbf{16.47} & \textbf{15.94} & \textbf{17.67} & \textbf{24.34} & \textbf{36.87} \\
          & \multicolumn{1}{c}{2} & & 26.60 & 26.62 & 27.07 & 27.49 & \textbf{27.91} & \textbf{28.08} & 27.36 & 25.77 & 23.98 & 22.39 & 20.29 & 19.87 & 23.39 & 37.65 & 67.28 \\
           \midrule
     & \multicolumn{1}{c}{0} & & 25.59 & 26.16 & 33.09 & 36.09 & 34.39 & 30.44 & 27.07 & 24.68 & 23.05 & 21.98 & 21.04 & 21.41 & 24.64 & 33.28 & 45.78 \\
     \textbf{0.25} & \multicolumn{1}{c}{1} & $Wi_c=$ & \textbf{21.66} & \textbf{21.77} & \textbf{23.32} & \textbf{24.91} & 26.95 & 26.77 & \textbf{24.41} & \textbf{21.61} & \textbf{19.30} & \textbf{17.67} & \textbf{16.04} & \textbf{15.80} & \textbf{17.71} & \textbf{24.27} & \textbf{36.23} \\
          & \multicolumn{1}{c}{2} & & 24.60 & 24.61 & 25.03 & 25.41 & \textbf{25.81} & \textbf{25.96} & 25.29 & 23.83 & 22.17 & 20.72 & 18.90 & 18.73 & 22.22 & 35.42 & 62.53 \\
           \midrule
     & \multicolumn{1}{c}{0} & & 25.68 & 26.24 & 33.09 & 36.09 & 34.39 & 30.44 & 27.09 & 24.75 & 23.18 & 22.17 & 21.36 & 21.84 & 25.28 & 34.26 & 47.18 \\
     \textbf{0.2} & \multicolumn{1}{c}{1} & $Wi_c=$ & \textbf{19.96} & \textbf{20.06} & \textbf{21.51} & \textbf{22.92} & 24.79 & 24.63 & \textbf{22.48} & \textbf{20.09} & \textbf{18.30} & \textbf{17.12} & \textbf{16.02} & \textbf{16.03} & \textbf{18.12} & \textbf{24.65} & \textbf{36.18} \\
          & \multicolumn{1}{c}{2} & & 22.67 & 22.68 & 23.07 & 23.42 & \textbf{23.78} & \textbf{23.92} & 23.30 & 21.95 & 20.46 & 19.19 & 17.83 & 17.95 & 21.43 & 33.61 & 58.25 \\
           \midrule
    & \multicolumn{1}{c}{0} & & 25.88 & 26.42 & 33.10 & 36.09 & 34.39 & 30.46 & 27.18 & 24.92 & 23.44 & 22.51 & 21.82 & 22.41 & 26.05 & 35.39 & 48.78 \\
        \textbf{0.167} & \multicolumn{1}{c}{1} & $Wi_c=$ & \textbf{18.97} & \textbf{19.09} & \textbf{20.50} & \textbf{21.68} & 23.30 & 23.20 & \textbf{21.33} & \textbf{19.41} & \textbf{18.03} & \textbf{17.13} & \textbf{16.32} & \textbf{16.49} & \textbf{18.72} & \textbf{25.33} & \textbf{36.67} \\
          & \multicolumn{1}{c}{2} & & 21.44 & 21.45 & 21.82 & 22.15 & \textbf{22.48} & \textbf{22.60} & 22.01 & 20.77 & 19.45 & 18.41 & 17.43 & 17.75 & 21.22 & 32.79 & 55.85 \\
           \midrule
   & \multicolumn{1}{c}{0} & & 26.24 & 26.76 & 33.16 & 36.10 & 34.41 & 30.55 & 27.40 & 25.27 & 23.88 & 23.03 & 22.46 & 23.17 & 27.04 & 36.81 & 50.78 \\
    \textbf{0.14} & \multicolumn{1}{c}{1} & $Wi_c=$ & \textbf{18.54} & \textbf{18.67} & \textbf{20.04} & \textbf{20.99} & 22.09 & 22.11 & \textbf{20.68} & \textbf{19.23} & \textbf{18.17} & \textbf{17.47} & \textbf{16.86} & \textbf{17.16} & \textbf{19.54} & \textbf{26.28} & \textbf{37.58} \\
          & \multicolumn{1}{c}{2} & & 20.45 & 20.47 & 20.83 & 21.14 & \textbf{21.44} & \textbf{21.51} & 20.98 & 19.91 & 18.85 & 18.06 & 17.40 & 17.87 & 21.34 & 32.46 & 54.28 \\
           \midrule
     & \multicolumn{1}{c}{0} & & 26.47 & 26.98 & 33.23 & 36.11 & 34.44 & 30.64 & 27.57 & 25.49 & 24.16 & 23.34 & 22.83 & 23.59 & 27.58 & 37.58 & 51.87 \\
     \textbf{0.129} & \multicolumn{1}{c}{1} & $Wi_c=$ & \textbf{18.52} & \textbf{18.65} & \textbf{20.00} & 20.87 & 21.77 & 21.78 & \textbf{20.57} & \textbf{19.30} & \textbf{18.35} & \textbf{17.72} & \textbf{17.19} & \textbf{17.54} & \textbf{19.99} & \textbf{26.83} & \textbf{38.15} \\
          & \multicolumn{1}{c}{2} & & 20.06 & 20.07 & 20.44 & \textbf{20.75} & \textbf{21.03} & \textbf{21.08} & 20.59 & 19.64 & 18.71 & 18.03 & 17.49 & 18.02 & 21.50 & 32.45 & 53.79 \\
           \midrule
     & \multicolumn{1}{c}{0} & & 27.45 & 27.94 & 33.66 & 36.29 & 34.69 & 31.18 & 28.39 & 26.52 & 25.32 & 24.60 & 24.25 & 25.19 & 29.59 & 40.42 & 55.83 \\
    \multirow{2}[0]{*}{\textbf{0.1}} & \multicolumn{1}{c}{1} & \multirow{2}[0]{*}{$Wi_c=$} & \textbf{19.04} & 19.18 & 20.48 & 21.21 & 21.76 & 21.70 & 20.93 & 20.04 & 19.33 & 18.84 & 18.50 & 18.99 & \textbf{21.71} & \textbf{28.93} & \textbf{40.49} \\
          & \multicolumn{1}{c}{2} & & 19.11 & \textbf{19.17} & \textbf{19.74} & \textbf{20.07} & \textbf{20.32} & \textbf{20.36} & \textbf{20.04} & \textbf{19.45} & \textbf{18.86} & \textbf{18.42} & \textbf{18.16} & \textbf{18.83} & 22.38 & 33.01 & 53.18 \\
          & \multicolumn{1}{c}{3} & & 24.07 & 24.07 & 24.33 & 24.54 & 24.66 & 24.54 & 24.01 & 23.11 & 22.16 & 21.41 & 20.84 & 21.68 & 26.85 & 43.71 &  \\
           \midrule
     & \multicolumn{1}{c}{0} & & 30.04 & 30.50 & 35.48 & 37.53 & 36.16 & 33.25 & 30.90 & 29.27 & 28.23 & 27.64 & 27.53 & 28.78 & 34.00 & 46.58 & 64.41 \\
    \multirow{2}[0]{*}{\textbf{0.066}} & \multicolumn{1}{c}{1} & \multirow{2}[0]{*}{$Wi_c=$} & 21.32 & 21.47 & 22.82 & 23.49 & 23.90 & 23.81 & 23.27 & 22.62 & 22.08 & 21.71 & 21.57 & 22.29 & 25.55 & \textbf{33.76} & \textbf{46.25} \\
          & \multicolumn{1}{c}{2} & & \textbf{19.63} & \textbf{19.72} & \textbf{20.56} & \textbf{20.97} & \textbf{21.28} & \textbf{21.39} & \textbf{21.25} & \textbf{20.94} & \textbf{20.61} & \textbf{20.37} & \textbf{20.38} & \textbf{21.23} & \textbf{25.03} & 35.67 & 54.86 \\
          & \multicolumn{1}{c}{3} & & 22.59 & 22.61 & 23.01 & 23.26 & 23.41 & 23.37 & 23.07 & 22.57 & 22.06 & 21.70 & 21.64 & 22.71 & 27.76 & 43.20 &  \\
           \midrule
     & \multicolumn{1}{c}{0} & & 33.61 & 34.07 & 38.68 & 40.30 & 39.14 & 36.68 & 34.58 & 33.08 & 32.12 & 31.60 & 31.68 & 33.25 & 39.42 & 54.11 & 74.87 \\
    \multirow{2}[0]{*}{\textbf{0.045}} & \multicolumn{1}{c}{1} & \multirow{2}[0]{*}{$Wi_c=$} & 24.73 & 24.90 & 26.43 & 27.14 & 27.52 & 27.41 & 26.91 & 26.31 & 25.81 & 25.49 & 25.49 & 26.44 & 30.38 & \textbf{39.93} & \textbf{53.92} \\
          & \multicolumn{1}{c}{2} & & \textbf{22.05} & \textbf{22.17} & \textbf{23.16} & \textbf{23.65} & \textbf{24.01} & \textbf{24.17} & \textbf{24.10} & \textbf{23.87} & \textbf{23.63} & \textbf{23.46} & \textbf{23.61} & \textbf{24.63} & \textbf{28.84} & 40.03 & 59.23 \\
          & \multicolumn{1}{c}{3} & & 22.69 & 22.78 & 23.61 & 24.01 & 24.30 & 24.43 & 24.35 & 24.13 & 23.88 & 23.71 & 23.91 & 25.14 & 30.31 & 45.25 & 73.52 \\
           \midrule
     & \multicolumn{1}{c}{0} & & 38.79 & 39.28 & 43.82 & 45.23 & 44.22 & 42.01 & 40.01 & 38.55 & 37.61 & 37.13 & 37.39 & 39.36 & 46.78 & 64.30 & 89.01 \\
     & \multicolumn{1}{c}{1} & & 29.72 & 29.93 & 31.75 & 32.56 & 32.96 & 32.80 & 32.25 & 31.61 & 31.11 & 30.80 & 30.92 & 32.17 & 37.07 & 48.58 & \textbf{64.92} \\
    \textbf{0.03} & \multicolumn{1}{c}{2} & $Wi_c=$ & 26.19 & 26.33 & 27.54 & 28.13 & 28.57 & 28.76 & 28.70 & 28.47 & 28.25 & 28.11 & 28.36 & 29.61 & \textbf{34.46} & \textbf{46.82} & 66.92 \\
          & \multicolumn{1}{c}{3} & & \textbf{25.42} & \textbf{25.54} & \textbf{26.60} & \textbf{27.12} & \textbf{27.52} & \textbf{27.76} & \textbf{27.79} & \textbf{27.68} & \textbf{27.54} & \textbf{27.46} & \textbf{27.83} & \textbf{29.24} & 34.78 & 49.99 & 77.42 \\
          & \multicolumn{1}{c}{4} & & 27.03 & 27.14 & 28.06 & 28.50 & 28.82 & 28.97 & 28.92 & 28.72 & 28.49 & 28.37 & 28.75 & 30.38 & 37.07 & 56.60 &  \\
           \midrule
     & \multicolumn{1}{c}{0} & & 61.89 & 62.55 & 68.39 & 69.97 & 69.00 & 66.60 & 64.25 & 62.45 & 61.32 & 60.81 & 61.62 & 65.11 & 77.64 & 106.92 & 148.10 \\
          & \multicolumn{1}{c}{1} & & 52.09 & 52.49 & 55.88 & 57.22 & 57.64 & 57.09 & 56.04 & 55.00 & 54.23 & 53.84 & 54.35 & 56.88 & 66.07 & 86.91 & 115.00 \\
     & \multicolumn{1}{c}{2} & & 46.40 & 46.68 & 49.11 & 50.20 & 50.89 & 51.02 & 50.70 & 50.19 & 49.75 & 49.53 & 50.07 & 52.31 & 60.52 & 79.75 & \textbf{107.61} \\
    \textbf{0.01} & \multicolumn{1}{c}{3} & $Wi_c=$ & 43.15 & 43.38 & 45.36 & 46.32 & 47.03 & 47.40 & 47.41 & 47.21 & 47.00 & 46.92 & 47.57 & 49.80 & 57.94 & \textbf{77.95} & 109.35 \\
          & \multicolumn{1}{c}{4} & & 41.46 & 41.66 & 43.44 & 44.33 & 45.03 & 45.48 & 45.64 & 45.60 & 45.51 & 45.53 & 46.31 & 48.64 & \textbf{57.19} & 79.22 &  \\
          & \multicolumn{1}{c}{5} & & \textbf{40.94} & \textbf{41.13} & \textbf{42.84} & \textbf{43.69} & \textbf{44.38} & \textbf{44.84} & \textbf{45.06} & \textbf{45.08} & \textbf{45.05} & \textbf{45.11} & \textbf{45.99} & \textbf{48.49} & 57.78 & 82.75 &  \\
          & \multicolumn{1}{c}{6} & & 41.52 & 41.71 & 43.40 & 44.22 & 44.88 & 45.32 & 45.51 & 45.51 & 45.47 & 45.53 & 46.12 & 49.19 & 59.45 & 88.29 &  \\
          \bottomrule
    \end{tabular}%
  \caption[Summarized results of the linear stability analysis of the isothermal, purely elastic Taylor-Couette base flow of various Oldroyd-B fluids in terms of the critical Weissenberg numbers]{Results of linear stability analysis of purely elastic ($Re=0$) Taylor-Couette base flow of an Oldroyd-B fluid. The table lists the critical Weissenberg numbers for the onset of elastic instability triggered by different perturbation modes $m$ in dependence of different gap widths $\varepsilon$ and viscosity ratios $\beta=\eta_s/\eta$. For each set $(\varepsilon,\beta)$, the smallest critical Weissenberg number (giving the most unstable perturbation mode) is highlighted in boldface.}
  \label{tab:summary_Wi_crit}%
\end{table}%

\end{landscape}

\bibliographystyle{elsarticle-num}
\bibliography{references}

\begin{thebibliography}{10}
\expandafter\ifx\csname url\endcsname\relax
  \def\url#1{\texttt{#1}}\fi
\expandafter\ifx\csname urlprefix\endcsname\relax\def\urlprefix{URL }\fi
\expandafter\ifx\csname href\endcsname\relax
  \def\href#1#2{#2} \def\path#1{#1}\fi

\bibitem{Chandrasekhar1961}
S.~Chandrasekhar, Hydrodynamic and Hydromagnetic Stability, Oxford Clarendon
  Press, 1961.

\bibitem{Larson1992}
R.~G. Larson, Instabilities in viscoelastic flows, Rheologica Acta 31~(3)
  (1992) 213--263.

\bibitem{Shaqfeh1996}
E.~S.~G. Shaqfeh, Purely elastic instabilities in viscometric flows, Annual
  Review of Fluid Mechanics 28~(1) (1996) 129--185.

\bibitem{Morozov2007}
A.~N. Morozov, W.~van Saarloos, An introductory essay on subcritical
  instabilities and the transition to turbulence in visco-elastic parallel
  shear flows, Physics Reports 447~(3--6) (2007) 112--143.

\bibitem{Groisman2004}
A.~Groisman, V.~Steinberg, Elastic turbulence in curvilinear flow of polymer
  solutions, New Journal of Physics 6~(29) (2004) 1--48.

\bibitem{Groisman2000}
A.~Groisman, V.~Steinberg, Elastic turbulence in a polymer solution flow,
  Nature 405 (2000) 53--55.

\bibitem{larson2000}
R.~G. Larson, Turbulence without inertia, Nature 405 (2000) 27.

\bibitem{Gorodtsov1967}
V.~Gorodtsov, A.~Leonov, On a linear instability of a plane parallele couette
  flow of viscoelastic fluid, Journal of Applied Mathematics and Mechanics 31
  (1967) 310--319.

\bibitem{Wilson1999}
H.~J. Wilson, M.~Renardy, Y.~Y. Renardy, {Structure of the spectrum in zero
  Reynolds number shear flow of the UCM and Oldroyd-B liquids}, {J. Non-Newton.
  Fluid Mech.} {80} ({1999}) {251--268}.

\bibitem{Arora2005}
K.~Arora, B.~Khomami, {The influence of finite extensibility on the
  eigenspectrum of dilute polymeric solutions}, {J. Non-Newton. Fluid Mech.}
  {129}~({1}) ({2005}) {56--60}.

\bibitem{Bertola2003}
V.~Bertola, B.~Meulenbroek, C.~Wagner, C.~Storm, A.~N. Morozov, W.~van
  Saarloos, D.~Bonn, Experimental evidence for an intrinsic route to polymer
  melt fracture phenomena: A nonlinear instability of viscoelastic poiseuille
  flow, Physical Review Letters 90~(11) (2003) 114502.
\newblock \href {http://dx.doi.org/10.1103/PhysRevLett.90.114502}
  {\path{doi:10.1103/PhysRevLett.90.114502}}.

\bibitem{Meulenbroek2004}
B.~Meulenbroek, C.~Storm, A.~N. Morozov, W.~van Saarloos, Weakly nonlinear
  subcritical instability of visco-elastic poiseuille flow, J. Non-Newton.
  Fluid Mech. 116~(2-3) (2004) 235--268.
\newblock \href {http://dx.doi.org/10.1016/j.jnnfm.2003.09.003}
  {\path{doi:10.1016/j.jnnfm.2003.09.003}}.

\bibitem{Morozov2005prl}
A.~Morozov, W.~van Saarloos, Subcritical {F}inite-{A}mplitude {S}olutions for
  {P}lane {C}ouette {F}low of {V}iscoelastic {F}luids, Phys. Rev. Lett.
  95~(024501).

\bibitem{Bonn2011}
D.~Bonn, F.~Ingremeau, Y.~Amarouchene, H.~Kellay, {Large velocity fluctuations
  in small-Reynolds-number pipe flow of polymer solutions}, {Phys. Rev. E}
  {84}~({4, 2}).
\newblock \href {http://dx.doi.org/{10.1103/PhysRevE.84.045301}}
  {\path{doi:{10.1103/PhysRevE.84.045301}}}.

\bibitem{Pan2013}
L.~Pan, A.~Morozov, C.~Wagner, P.~E. Arratia, {Nonlinear Elastic Instability in
  Channel Flows at Low Reynolds Numbers}, {Phys. Rev. Lett.} {110}~({17}).
\newblock \href {http://dx.doi.org/{10.1103/PhysRevLett.110.174502}}
  {\path{doi:{10.1103/PhysRevLett.110.174502}}}.

\bibitem{Pakdel1996}
P.~Pakdel, G.~H. McKinley, Elastic instability and curved streamlines, Physical
  Review Letters 77~(12) (1996) 2459--2462.

\bibitem{McKinley1996}
G.~H. McKinley, P.~Pakdel, A.~Oeztekin, Rheological and geometric scaling of
  purely elastic flow instabilities, J. Non-Newton. Fluid Mech. 67 (1996)
  19--47.

\bibitem{Muller1989}
S.~J. Muller, R.~G. Larson, E.~S.~G. Shaqfeh, A purely elastic transition in
  {T}aylor-{C}ouette flow, Rheologica Acta 28~(6) (1989) 499--503.

\bibitem{Larson1990}
R.~G. Larson, E.~S.~G. Shaqfeh, S.~J. Muller, A purely elastic instability in
  {T}aylor-{C}ouette flow, Journal of Fluid Mechanics 218 (1990) 573--600.

\bibitem{Larson1994}
R.~G. Larson, S.~J. Muller, E.~S.~G. Shaqfeh, {The effect of fluid rheology on
  the elastic Taylor-Couette instability}, J. Non-Newton. Fluid Mech. 51~(2)
  (1994) 195--225.

\bibitem{Weissenberg1948}
K.~Weissenberg, in: Proceedings of the First International Rheological
  Congress, Vol.~1, 1948.

\bibitem{Groisman1996}
A.~Groisman, V.~Steinberg, {C}ouette-{T}aylor flow in a dilute polymer
  solution, Physical Review Letters 77~(8) (1996) 1480--1483.

\bibitem{Groisman1998}
A.~Groisman, V.~Steinberg, Mechanism of elastic instability in {C}ouette flow
  of polymer solutions: {E}xperiment, Physics of Fluids 10~(10) (1998)
  2451--2463.

\bibitem{Groisman1998a}
A.~Groisman, V.~Steinberg, Elastic vs. inertial instability in a polymer
  solution flow, Europhysics Letters 43~(2) (1998) 165--170.

\bibitem{Zilz2012}
J.~Zilz, R.~J. Poole, M.~A. Alves, D.~Bartolo, B.~Levaché, A.~Lindner,
  Geometric scaling of purely elastic flow instability in serpentine channels,
  Journal of Fluid Mechanics 712 (2012) 203--218.

\bibitem{Poole2013}
R.~J. Poole, A.~Lindner, M.~A. Alves, {Viscoelastic secondary flows in
  serpentine channels}, {J. Non-Newton. Fluid Mech.} {201} ({2013}) {10--16}.
\newblock \href {http://dx.doi.org/{10.1016/j.jnnfm.2013.07.001}}
  {\path{doi:{10.1016/j.jnnfm.2013.07.001}}}.

\bibitem{Alves2007}
M.~A. Alves, R.~J. Poole, {Divergent flow in contractions}, {J. Non-Newton.
  Fluid Mech.} {144}~({2-3}) ({2007}) {140--148}.
\newblock \href {http://dx.doi.org/{10.1016/j.jnnfm.2007.04.003}}
  {\path{doi:{10.1016/j.jnnfm.2007.04.003}}}.

\bibitem{Giesekus1966}
H.~Giesekus, {Zur Stabilitaet von Stroemungen viskoelastischer Fluessigkeiten},
  Rheologica Acta 5~(3) (1966) 239--252.

\bibitem{Shaqfeh1992}
E.~S.~G. Shaqfeh, S.~J. Muller, R.~G. Larson, The effects of gap width and
  dilute solution properties on the viscoelastic {T}aylor-{C}ouette
  instability, Journal of Fluid Mechanics 235 (1992) 285--317.

\bibitem{Avgousti1993}
M.~Avgousti, A.~N. Beris, Non-axisymmetric modes in viscoelastic
  {T}aylor-{C}ouette flow, J. Non-Newton. Fluid Mech. 50~(2--3) (1993)
  225--251.

\bibitem{Avgousti1993a}
M.~Avgousti, A.~N. Beris, Viscoelastic {T}aylor-{C}ouette flow: {B}ifurcation
  analysis in the presence of symmetries, Proceedings of the Royal Society of
  London A - Mathematical and Physical Sciences 443~(1917) (1993) 17--37.

\bibitem{Joo1994}
Y.~L. Joo, E.~S.~G. Shaqfeh, Observations of purely elastic instabilities in
  the {T}aylor-{D}ean flow of a {B}oger fluid, Journal of Fluid Mechanics 262
  (1994) 27--73.

\bibitem{Thomas2006}
D.~G. Thomas, R.~Sureshkumar, B.~Khomami, {Pattern formation in Taylor-Couette
  flow of dilute polymer solutions: Dynamical simulations and mechanism},
  {Phys. Rev. Lett.} {97}~({5}) ({2006}) {054501}.
\newblock \href {http://dx.doi.org/{10.1103/PhysRevLett.97.054501}}
  {\path{doi:{10.1103/PhysRevLett.97.054501}}}.

\bibitem{Thomas2006a}
D.~G. Thomas, U.~A. Al-Mubaiyedh, R.~Sureshkumar, B.~Khomami, {Time-dependent
  simulations of non-axisymmetric patterns in Taylor-Couette flow of dilute
  polymer solutions}, {Journal of Non-Newtonian Fluid Mechanics} {138}~({2-3})
  ({2006}) {111--133}.
\newblock \href {http://dx.doi.org/{10.1016/j.jnnfm.2006.04.013}}
  {\path{doi:{10.1016/j.jnnfm.2006.04.013}}}.

\bibitem{Thomas2009}
D.~G. Thomas, B.~Khomami, R.~Sureshkumar, {Nonlinear dynamics of viscoelastic
  Taylor-Couette flow: effect of elasticity on pattern selection, molecular
  conformation and drag}, {J. Fluid Mech.} {620} ({2009}) {353--382}.
\newblock \href {http://dx.doi.org/{10.1017/S0022112008004710}}
  {\path{doi:{10.1017/S0022112008004710}}}.

\bibitem{Khomami2013}
N.~Liu, B.~Khomami, {Elastically induced turbulence in Taylor-Couette flow:
  direct numerical simulation and mechanistic insight}, {J. Fluid Mech.} {737}
  ({2013}) {R4}.
\newblock \href {http://dx.doi.org/{10.1017/jfm.2013.544}}
  {\path{doi:{10.1017/jfm.2013.544}}}.

\bibitem{Baumert1999}
B.~M. Baumert, S.~J. Muller, Axisymmetric and non-axisymmetric elastic and
  inertio-elastic instabilities in {T}aylor-{C}ouette flow, J. Non-Newton.
  Fluid Mech. 83~(1--2) (1999) 33--69.

\bibitem{Crumeyrolle2002}
O.~Crumeyrolle, I.~Mutabazi, M.~Grisel, Experimental study of inertioelastic
  {C}ouette-{T}aylor instability modes in dilute and semidilute polymer
  solutions, Physics of Fluids 14~(5) (2002) 1681--1688.

\bibitem{Dutcher2013}
C.~S. Dutcher, S.~J. Muller, {Effects of moderate elasticity on the stability
  of co- and counter-rotating Taylor-Couette flows}, {JOURNAL OF RHEOLOGY}
  {57}~({3}) ({2013}) {791--812}.
\newblock \href {http://dx.doi.org/{10.1122/1.4798549}}
  {\path{doi:{10.1122/1.4798549}}}.

\bibitem{Fardin2014}
M.~A. Fardin, C.~Perge, N.~Taberlet, {"}the hydrogen atom of fluid dynamics{"}
  - introduction to the taylor-couette flow for soft matter scientists, Soft
  Matter 10 (2014) 3523--3535.
\newblock \href {http://dx.doi.org/10.1039/C3SM52828F}
  {\path{doi:10.1039/C3SM52828F}}.

\bibitem{Becu:2007}
L.~{Becu \emph{et al.}}, Evidence for three-dimensional unstable flows in
  shear-banding wormlike micelles, Phys. Rev. E 76 (2007) 011503.

\bibitem{Fardin:2009}
M.~A. {Fardin \emph{et al.}}, Taylor-like vortices in shear-banding flow of
  giant micelles, Phys. Rev. Lett. 103 (2009) 028302.

\bibitem{Fardin:2010}
M.~A. {Fardin \emph{et al.}}, Elastic turbulence in shear banding wormlike
  micelles, Phys. Rev. Lett. 104 (2010) 178303.

\bibitem{Decruppe:2010}
J.~P. {Decruppe \emph{et al.}}, Azimuthal instability of the interface in a
  shear banded flow by direct visual observation, Phys. Rev. Lett. 105 (2010)
  258301.

\bibitem{Fielding:2010}
S.~M. Fielding, Viscoelastic taylor-couette instability of shear banded flow,
  Phys. Rev. Lett. 104 (2010) 198303.

\bibitem{Nicolas2012}
A.~Nicolas, A.~Morozov, {Nonaxisymmetric Instability of Shear-Banded
  Taylor-Couette Flow}, {Phys. Rev. Lett.} {108}~({8}) ({2012}) {088302}.
\newblock \href {http://dx.doi.org/{10.1103/PhysRevLett.108.088302}}
  {\path{doi:{10.1103/PhysRevLett.108.088302}}}.

\bibitem{Mohammadigoushki2017}
H.~Mohammadigoushki, S.~J. Muller, {Inertio-elastic instability in
  Taylor-Couette flow of a model wormlike micellar system}, {JOURNAL OF
  RHEOLOGY} {61}~({4}) ({2017}) {683--696}.
\newblock \href {http://dx.doi.org/{10.1122/1.4983843}}
  {\path{doi:{10.1122/1.4983843}}}.

\bibitem{Nicolas2016}
A.~Nicolas, M.~Fuchs, {Shear-thinning in dense colloidal suspensions and its
  effect on elastic instabilities: From the microscopic equations of motion to
  an approximation of the macroscopic rheology}, {J. Non-Newton. Fluid Mech.}
  {228} ({2016}) {64--78}.
\newblock \href {http://dx.doi.org/{10.1016/j.jnnfm.2015.12.010}}
  {\path{doi:{10.1016/j.jnnfm.2015.12.010}}}.

\bibitem{PhanThien1977}
N.~Phan-Thien, R.~I. Tanner, A new constitutive equation derived from network
  theory, J. Non-Newton. Fluid Mech. 2~(4) (1977) 353--365.

\bibitem{Bird1987}
R.~B. Bird, R.~C. Armstrong, O.~Hassager, Dynamics of polymeric liquids: Fluid
  mechanics, 2nd Edition, Vol.~1, John Wiley and Sons Inc., New York, NY, 1987.

\bibitem{Bird1995}
R.~B. Bird, J.~M. Wiest, {Constitutive Equations for Polymeric Liquids}, Annual
  Review of Fluid Mechanics 27 (1995) 169--193.

\bibitem{Mirzazadeh2005}
M.~Mirzazadeh, M.~P. Escudier, F.~Rashidi, S.~H. Hashemabadi, Purely tangential
  flow of a {PTT}-viscoelastic fluid within a concentric annulus, J.
  Non-Newton. Fluid Mech. 129~(2) (2005) 88--97.

\bibitem{Zell2010}
A.~Zell, S.~E. Gier, S.~Rafa\"{i}, C.~Wagner, {Is there a relation between the
  relaxation time measured in CaBER experiments and the first normal stress
  coefficient?}, J. Non-Newton. Fluid Mech. 165~(19--20) (2010) 1265--1274.

\bibitem{Groisman2001}
A.~Groisman, V.~Steinberg, {Efficient mixing at low Reynolds numbers using
  polymer additives}, Nature 410 (2001) 905--907.

\bibitem{Elbing2009}
B.~R. Elbing, E.~S. Winkel, M.~J. Solomon, S.~L. Ceccio, Degradation of
  homogeneous polymer solutions in high shear turbulent pipe flow, Experiments
  in Fluids 47~(6) (2009) 1033--1044.

\bibitem{Owolabi2017}
B.~E. Owolabi, D.~J.~C. Dennis, R.~J. Poole, {Turbulent drag reduction by
  polymer additives in parallel-shear flows}, {J. Fluid Mech.} {827} ({2017})
  {R4}.
\newblock \href {http://dx.doi.org/{10.1017/jfm.2017.544}}
  {\path{doi:{10.1017/jfm.2017.544}}}.

\bibitem{Casanellas2016}
L.~Casanellas, M.~A. Alves, R.~J. Poole, S.~Lerouge, A.~Lindner, {The
  stabilizing effect of shear thinning on the onset of purely elastic
  instabilities in serpentine microflows}, {Soft Matter} {12}~({29}) ({2016})
  {6167--6175}.
\newblock \href {http://dx.doi.org/{10.1039/c6sm00326e}}
  {\path{doi:{10.1039/c6sm00326e}}}.

\bibitem{Esser1996}
A.~Esser, S.~Grossmann, Analytic expression for {T}aylor-{C}ouette stability
  boundary, Physics of Fluids 8~(7) (1996) 1814--1819.

\bibitem{Dutcher2007}
C.~S. Dutcher, S.~J. Muller, Explicit analytic formulas for {N}ewtonian
  {T}aylor-{C}ouette primary instabilities, Physical Review E 75~(4) (2007)
  047301.

\bibitem{Al-Mubaiyedh1999}
U.~A. Al-Mubaiyedh, R.~Sureshkumar, B.~Khomami, Influence of energetics on the
  stability of viscoelastic {T}aylor-{C}ouette flow, Physics of Fluids 11~(11)
  (1999) 3217--3226.

\bibitem{Al-Mubaiyedh2000}
U.~A. Al-Mubaiyedh, R.~Sureshkumar, B.~Khomami, Linear stability of
  viscoelastic {T}aylor-{C}ouette flow: {I}nfluence of fluid rheology and
  energetics, Journal of Rheology 44~(5) (2000) 1121--1138.

\bibitem{Canuto:book}
C.~Canuto, M.~Hussaini, A.~Quarteroni, T.~Zang, Spectral Methods in Fluid
  Dynamics, Springer Verlag, 1988.

\bibitem{scipy}
E.~Jones, T.~Oliphant, P.~Peterson, et~al., {SciPy}: Open source scientific
  tools for {Python}, http://www.scipy.org/ (2001--).

\bibitem{Faisst2000}
H.~Faisst, B.~Eckhardt, Transition from the {C}ouette-{T}aylor system to the
  plane {C}ouette system, Physical Review E 61~(6) (2000) 7227--7230.

\bibitem{White2000}
J.~M. White, S.~J. Muller, Viscous heating and the stability of {N}ewtonian and
  viscoelastic {T}aylor-{C}ouette flows, Physical Review Letters 84~(22) (2000)
  5130--5133.

\bibitem{White2003}
J.~M. White, S.~J. Muller, Experimental studies on the effect of viscous
  heating on the hydrodynamic stability of viscoelastic {T}aylor-{C}ouette
  flow, Journal of Rheology 47~(6) (2003) 1467--1492.

\bibitem{James2009}
D.~James, Boger fluids, Annual Review of Fluid Mechanics 41 (2009) 129--142.
\newblock \href {http://dx.doi.org/10.1146/annurev.fluid.010908.165125}
  {\path{doi:10.1146/annurev.fluid.010908.165125}}.

\bibitem{bird2}
R.~B. Bird, C.~F. Curtiss, R.~C. Armstrong, O.~Hassager, Dynamics of polymeric
  liquids, 2nd Edition, Vol. 2. Kinetic theory, Wiley, New York, 1987.

\bibitem{Morozov2015}
A.~N. Morozov, S.~E. Spagnolie, {Introduction to Complex Fluids}, in: S.~E.
  Spagnolie (Ed.), Complex Fluids in Biological Systems, Biological and Medical
  Physics, Biomedical Engineering, Springer New York, New York, NY, 2015, pp.
  3--52.
\newblock \href {http://dx.doi.org/10.1007/978-1-4939-2065-5}
  {\path{doi:10.1007/978-1-4939-2065-5}}.

\bibitem{DoiEdwards}
M.~Doi, S.~F. Edwards, The Theory of Polymer Dynamics, Oxford University Press,
  1986.

\bibitem{Tanner1985}
R.~I. Tanner, Engeneering Rheology, Oxford Science, 1985.

\bibitem{ll87}
L.~D. Landau, E.~M. Lifshitz, {Fluid Mechanics, Vol. 6}, translated from
  Russian by J. B. Sykes and W. H. Reid; Butterworth-Heinemann, Oxford, UK,
  1987.

\bibitem{Schafer2013}
C.~Sch\"{a}fer,
  \href{https://publikationen.sulb.uni-saarland.de/handle/20.500.11880/23017;jsessionid=4A53C64F40B707412D83441BF0451A12}{Elastic
  flow instabilities of non-newtonian fluids in shear flows}, Ph.D. thesis,
  Universit\"{a}t des Saarlandes (2013).
\newline\urlprefix\url{https://publikationen.sulb.uni-saarland.de/handle/20.500.11880/23017;jsessionid=4A53C64F40B707412D83441BF0451A12}

\bibitem{Broniarz-Press2009}
L.~Broniarz-Press, K.~Pralat, Thermal conductivity of newtonian and
  non-newtonian liquids, International Journal of Heat and Mass Transfer
  52~(21--22) (2009) 4701--4710.

\bibitem{Cheng2008}
N.~Cheng, Formula for the viscosity of a glycerol-water mixture, Industrial \&
  Engineering Chemistry Research 47~(9) (2008) 3285--3288.
\newblock \href {http://dx.doi.org/10.1021/ie071349z}
  {\path{doi:10.1021/ie071349z}}.

\bibitem{Traore2015}
B.~Traore, C.~Castelain, T.~Burghelea, {Efficient heat transfer in a regime of
  elastic turbulence}, {J. Non-Newton. Fluid Mech.} {223} ({2015}) {62--76}.
\newblock \href {http://dx.doi.org/{10.1016/j.jnnfm.2015.05.005}}
  {\path{doi:{10.1016/j.jnnfm.2015.05.005}}}.

\bibitem{Abed2016}
W.~M. Abed, R.~D. Whalley, D.~J.~C. Dennis, R.~J. Poole, {Experimental
  investigation of the impact of elastic turbulence on heat transfer in a
  serpentine channel}, {J. Non-Newton. Fluid Mech.} {231} ({2016}) {68--78}.
\newblock \href {http://dx.doi.org/{10.1016/j.jnnfm.2016.03.003}}
  {\path{doi:{10.1016/j.jnnfm.2016.03.003}}}.

\end{thebibliography}

\end{document}